\documentclass[runningheads,envcountsame]{llncs}

\newif\iflongversion
\longversiontrue

\usepackage[T1]{fontenc}
\usepackage{graphicx}
\usepackage[colorlinks]{hyperref}
\hypersetup{
	colorlinks=true,
	linkcolor=blue,
	filecolor=magenta,
	urlcolor=blue,
	citecolor = blue,
}
\usepackage{color}

\usepackage{paralist}
\usepackage{xspace}
\usepackage{listings}

\usepackage{bbding}
\usepackage{prettyref}
\newcommand{\rref}[2][]{\prettyref{#2}}
\newrefformat{sec}{Section\,\ref{#1}}
\newrefformat{subsec}{Section\,\ref{#1}}
\newrefformat{app}{Appendix\,\ref{#1}}
\newrefformat{def}{Def.\,\ref{#1}}
\newrefformat{thm}{Theorem\,\ref{#1}}
\newrefformat{prop}{Proposition\,\ref{#1}}
\newrefformat{lem}{Lemma\,\ref{#1}}
\newrefformat{cor}{Corollary\,\ref{#1}}
\newrefformat{rem}{Remark\,\ref{#1}}
\newrefformat{ex}{Example\,\ref{#1}}
\newrefformat{tab}{Table\,\ref{#1}}
\newrefformat{fig}{Fig.\,\ref{#1}}
\newrefformat{case}{case\,\ref{#1}}
\newrefformat{foot}{Footnote\,\ref{#1}}
\newrefformat{itm}{\ref{#1}}
\newrefformat{item}{\ref{#1}}
\newrefformat{alg}{\ref{#1}}

\newcommand{\isabellehol}{Isabelle/HOL\xspace}

\usepackage{tikz}

\usetikzlibrary{arrows}
\usetikzlibrary{positioning}
\usetikzlibrary{fit}

\usepackage{pgfplots}
\usepackage{pgf-pie}
\usetikzlibrary{arrows,automata}
\usetikzlibrary{fadings,shapes.arrows,shadows}
\tikzfading[name=arrowfading, top color=transparent!0, bottom color=transparent!100]
\tikzset{arrowfill/.style={top color=red!10, bottom color=red, general shadow={fill=black, shadow yshift=-0.8ex, path fading=arrowfading}}}
\tikzset{arrowstyle/.style={draw=gray,arrowfill, single arrow,minimum height=#1, single arrow,
single arrow head extend=.4cm,}}

\usetikzlibrary{plotmarks}

\makeatletter
\tikzset{
    scale plot marks/.is choice,
    scale plot marks/false/.code={
        \def\pgfuseplotmark##1{\pgftransformresetnontranslations\csname pgf@plot@mark@##1\endcsname}
    },
    scale plot marks/true/.style={},
    scale plot marks/.default=true
}
\makeatother

\usepackage{graphicx}
\usepackage{xcolor,mdframed}
\usepackage{fancyvrb}
\usepackage{isabelle}
\usepackage{isabellesym}
\newcommand{\snip}[4]{\expandafter\newcommand\csname #1\endcsname{#4}}
\snip{medianboundraw}{1}{2}{%
\isacommand{lemma}\isamarkupfalse%
\ median{\isacharunderscore}{\kern0pt}bound{\isacharunderscore}{\kern0pt}raw{\isacharcolon}{\kern0pt}\isanewline
\isakeyword{assumes}\ {\isachardoublequoteopen}prob{\isacharunderscore}{\kern0pt}space{\isachardot}{\kern0pt}indep{\isacharunderscore}{\kern0pt}vars\ {\isacharparenleft}{\kern0pt}measure{\isacharunderscore}{\kern0pt}pmf\ p{\isacharparenright}{\kern0pt}\ {\isacharparenleft}{\kern0pt}{\isasymlambda}{\isacharunderscore}{\kern0pt}{\isachardot}{\kern0pt}\ borel{\isacharparenright}{\kern0pt}\ Ys\ {\isacharbraceleft}{\kern0pt}{\isadigit{0}}{\isachardot}{\kern0pt}{\isachardot}{\kern0pt}{\isacharless}{\kern0pt}t{\isacharbraceright}{\kern0pt}{\isachardoublequoteclose}\isanewline
\isakeyword{assumes}\ {\isachardoublequoteopen}{\isasymAnd}i{\isachardot}{\kern0pt}\ i\ {\isacharless}{\kern0pt}\ t\ {\isasymLongrightarrow}\ measure{\isacharunderscore}{\kern0pt}pmf{\isachardot}{\kern0pt}prob\ p\ {\isacharbraceleft}{\kern0pt}c{\isachardot}{\kern0pt}\ Ys\ i\ c\ {\isasymin}\ {\isacharbraceleft}{\kern0pt}a{\isachardot}{\kern0pt}{\isachardot}{\kern0pt}b{\isacharbraceright}{\kern0pt}{\isacharbraceright}{\kern0pt}\ {\isacharequal}{\kern0pt}\ r{\isachardoublequoteclose}\isanewline
\isakeyword{assumes}\ {\isachardoublequoteopen}{\isadigit{0}}\ {\isasymle}\ r{\isachardoublequoteclose}\ {\isachardoublequoteopen}r\ {\isasymle}\ {\isadigit{1}}{\isachardoublequoteclose}\isanewline
\isakeyword{shows}\ {\isachardoublequoteopen}\isanewline
\ \ measure{\isacharunderscore}{\kern0pt}pmf{\isachardot}{\kern0pt}prob\ p\ {\isacharbraceleft}{\kern0pt}c{\isachardot}{\kern0pt}\ median\ t\ {\isacharparenleft}{\kern0pt}{\isasymlambda}i{\isachardot}{\kern0pt}\ Ys\ i\ c{\isacharparenright}{\kern0pt}\ {\isasymin}\ {\isacharbraceleft}{\kern0pt}a{\isachardot}{\kern0pt}{\isachardot}{\kern0pt}b{\isacharcolon}{\kern0pt}{\isacharcolon}{\kern0pt}real{\isacharbraceright}{\kern0pt}{\isacharbraceright}{\kern0pt}\isanewline
\ \ \ \ {\isasymge}\ {\isadigit{1}}\ {\isacharminus}{\kern0pt}\ measure{\isacharunderscore}{\kern0pt}pmf{\isachardot}{\kern0pt}prob\ {\isacharparenleft}{\kern0pt}binomial{\isacharunderscore}{\kern0pt}pmf\ t\ r{\isacharparenright}{\kern0pt}\ {\isacharbraceleft}{\kern0pt}{\isachardot}{\kern0pt}{\isachardot}{\kern0pt}t\ div\ {\isadigit{2}}{\isacharbraceright}{\kern0pt}{\isachardoublequoteclose}%
}
\snip{probbinomial}{1}{2}{%
\isacommand{lemma}\isamarkupfalse%
\ prob{\isacharunderscore}{\kern0pt}binomial{\isacharunderscore}{\kern0pt}pmf{\isacharunderscore}{\kern0pt}upto{\isacharcolon}{\kern0pt}\isanewline
\isakeyword{assumes}\ {\isachardoublequoteopen}{\isadigit{0}}\ {\isasymle}\ r{\isachardoublequoteclose}\ {\isachardoublequoteopen}r\ {\isasymle}\ {\isadigit{1}}{\isachardoublequoteclose}\isanewline
\isakeyword{shows}\ {\isachardoublequoteopen}\isanewline
\ \ measure{\isacharunderscore}{\kern0pt}pmf{\isachardot}{\kern0pt}prob\ {\isacharparenleft}{\kern0pt}binomial{\isacharunderscore}{\kern0pt}pmf\ t\ r{\isacharparenright}{\kern0pt}\ {\isacharbraceleft}{\kern0pt}{\isachardot}{\kern0pt}{\isachardot}{\kern0pt}m{\isacharbraceright}{\kern0pt}\ {\isacharequal}{\kern0pt}\isanewline
\ \ \ \ {\isacharparenleft}{\kern0pt}{\isasymSum}i\ {\isacharequal}{\kern0pt}\ {\isadigit{0}}{\isachardot}{\kern0pt}{\isachardot}{\kern0pt}m{\isachardot}{\kern0pt}\ {\isacharparenleft}{\kern0pt}t\ choose\ i{\isacharparenright}{\kern0pt}\ {\isacharasterisk}{\kern0pt}\ r\ {\isacharcircum}{\kern0pt}\ i\ {\isacharasterisk}{\kern0pt}\ \ {\isacharparenleft}{\kern0pt}{\isadigit{1}}\ {\isacharminus}{\kern0pt}\ r{\isacharparenright}{\kern0pt}\ {\isacharcircum}{\kern0pt}\ {\isacharparenleft}{\kern0pt}t\ {\isacharminus}{\kern0pt}\ i{\isacharparenright}{\kern0pt}{\isacharparenright}{\kern0pt}{\isachardoublequoteclose}%
}
\snip{appmclcertsound}{1}{2}{%
\isacommand{theorem}\isamarkupfalse%
\ certcheck{\isacharunderscore}{\kern0pt}sound{\isacharcolon}{\kern0pt}\isanewline
\isakeyword{assumes}\ {\isachardoublequoteopen}{\isacharparenleft}{\kern0pt}{\isasymAnd}F{\isachardot}{\kern0pt}\ check{\isacharunderscore}{\kern0pt}unsat\ F\ {\isasymLongrightarrow}\ sols\ F\ {\isacharequal}{\kern0pt}\ {\isacharbraceleft}{\kern0pt}{\isacharbraceright}{\kern0pt}{\isacharparenright}{\kern0pt}{\isachardoublequoteclose}\isanewline
\isakeyword{assumes}\ {\isachardoublequoteopen}{\isasymdelta}\ {\isachargreater}{\kern0pt}\ {\isadigit{0}}{\isachardoublequoteclose}\ {\isachardoublequoteopen}{\isasymdelta}\ {\isacharless}{\kern0pt}\ {\isadigit{1}}{\isachardoublequoteclose}\ {\isachardoublequoteopen}{\isasymepsilon}\ {\isachargreater}{\kern0pt}\ {\isadigit{0}}{\isachardoublequoteclose}\ {\isachardoublequoteopen}distinct\ S{\isachardoublequoteclose}\isanewline
\isakeyword{shows}\ {\isachardoublequoteopen}let\ sz\ {\isacharequal}{\kern0pt}\ real\ {\isacharparenleft}{\kern0pt}card\ {\isacharparenleft}{\kern0pt}proj\ {\isacharparenleft}{\kern0pt}set\ S{\isacharparenright}{\kern0pt}\ {\isacharparenleft}{\kern0pt}sols\ F{\isacharparenright}{\kern0pt}{\isacharparenright}{\kern0pt}{\isacharparenright}{\kern0pt}\ in\ \isanewline
\ \ measure{\isacharunderscore}{\kern0pt}pmf{\isachardot}{\kern0pt}prob\isanewline
\ \ \ \ {\isacharparenleft}{\kern0pt}map{\isacharunderscore}{\kern0pt}pmf\ {\isacharparenleft}{\kern0pt}{\isasymlambda}r{\isachardot}{\kern0pt}\ certcheck\ check{\isacharunderscore}{\kern0pt}unsat\ F\ S\ {\isasymepsilon}\ {\isasymdelta}\ {\isacharparenleft}{\kern0pt}f\ r{\isacharparenright}{\kern0pt}\ r{\isacharparenright}{\kern0pt}\isanewline
\ \ \ \ \ \ {\isacharparenleft}{\kern0pt}random{\isacharunderscore}{\kern0pt}seed{\isacharunderscore}{\kern0pt}xors\ {\isacharparenleft}{\kern0pt}find{\isacharunderscore}{\kern0pt}t\ {\isasymdelta}{\isacharparenright}{\kern0pt}\ {\isacharparenleft}{\kern0pt}length\ S{\isacharparenright}{\kern0pt}{\isacharparenright}{\kern0pt}{\isacharparenright}{\kern0pt}\isanewline
\ \ \ \ {\isacharbraceleft}{\kern0pt}c{\isachardot}{\kern0pt}\ {\isasymnot}isl\ c\ {\isasymand}\ projr\ c\ {\isasymnotin}\ {\isacharbraceleft}{\kern0pt}sz\ {\isacharslash}{\kern0pt}\ {\isacharparenleft}{\kern0pt}{\isadigit{1}}\ {\isacharplus}{\kern0pt}\ {\isasymepsilon}{\isacharparenright}{\kern0pt}\ {\isachardot}{\kern0pt}{\isachardot}{\kern0pt}\ {\isacharparenleft}{\kern0pt}{\isadigit{1}}\ {\isacharplus}{\kern0pt}\ {\isasymepsilon}{\isacharparenright}{\kern0pt}\ {\isacharasterisk}{\kern0pt}\ sz{\isacharbraceright}{\kern0pt}{\isacharbraceright}{\kern0pt}\ {\isasymle}\ {\isasymdelta}{\isachardoublequoteclose}%
}
\snip{appmclcertcomplete}{1}{2}{%
\isacommand{theorem}\isamarkupfalse%
\ certcheck{\isacharunderscore}{\kern0pt}promise{\isacharunderscore}{\kern0pt}complete{\isacharcolon}{\kern0pt}\isanewline
\isakeyword{assumes}\ {\isachardoublequoteopen}{\isacharparenleft}{\kern0pt}{\isasymAnd}F{\isachardot}{\kern0pt}\ check{\isacharunderscore}{\kern0pt}unsat\ F\ {\isasymLongrightarrow}\ sols\ F\ {\isacharequal}{\kern0pt}\ {\isacharbraceleft}{\kern0pt}{\isacharbraceright}{\kern0pt}{\isacharparenright}{\kern0pt}{\isachardoublequoteclose}\isanewline
\isakeyword{assumes}\ {\isachardoublequoteopen}{\isasymdelta}\ {\isachargreater}{\kern0pt}\ {\isadigit{0}}{\isachardoublequoteclose}\ {\isachardoublequoteopen}{\isasymdelta}\ {\isacharless}{\kern0pt}\ {\isadigit{1}}{\isachardoublequoteclose}\ {\isachardoublequoteopen}{\isasymepsilon}\ {\isachargreater}{\kern0pt}\ {\isadigit{0}}{\isachardoublequoteclose}\ {\isachardoublequoteopen}distinct\ S{\isachardoublequoteclose}\isanewline
\isakeyword{assumes}\ {\isachardoublequoteopen}{\isasymAnd}r{\isachardot}{\kern0pt}\ r\ {\isasymin}\ set{\isacharunderscore}{\kern0pt}pmf\ {\isacharparenleft}{\kern0pt}random{\isacharunderscore}{\kern0pt}seed{\isacharunderscore}{\kern0pt}xors\ {\isacharparenleft}{\kern0pt}find{\isacharunderscore}{\kern0pt}t\ {\isasymdelta}{\isacharparenright}{\kern0pt}\ {\isacharparenleft}{\kern0pt}length\ S{\isacharparenright}{\kern0pt}{\isacharparenright}{\kern0pt}{\isasymLongrightarrow}\isanewline
\ \ {\isasymnot}isl\ {\isacharparenleft}{\kern0pt}certcheck\ check{\isacharunderscore}{\kern0pt}unsat\ F\ S\ {\isasymepsilon}\ {\isasymdelta}\ {\isacharparenleft}{\kern0pt}f\ r{\isacharparenright}{\kern0pt}\ r{\isacharparenright}{\kern0pt}{\isachardoublequoteclose}\isanewline
\isakeyword{shows}\ {\isachardoublequoteopen}let\ sz\ {\isacharequal}{\kern0pt}\ real\ {\isacharparenleft}{\kern0pt}card\ {\isacharparenleft}{\kern0pt}proj\ {\isacharparenleft}{\kern0pt}set\ S{\isacharparenright}{\kern0pt}\ {\isacharparenleft}{\kern0pt}sols\ F{\isacharparenright}{\kern0pt}{\isacharparenright}{\kern0pt}{\isacharparenright}{\kern0pt}\ in\ \isanewline
\ \ measure{\isacharunderscore}{\kern0pt}pmf{\isachardot}{\kern0pt}prob\isanewline
\ \ \ \ {\isacharparenleft}{\kern0pt}map{\isacharunderscore}{\kern0pt}pmf\ {\isacharparenleft}{\kern0pt}{\isasymlambda}r{\isachardot}{\kern0pt}\ certcheck\ check{\isacharunderscore}{\kern0pt}unsat\ F\ S\ {\isasymepsilon}\ {\isasymdelta}\ {\isacharparenleft}{\kern0pt}f\ r{\isacharparenright}{\kern0pt}\ r{\isacharparenright}{\kern0pt}\isanewline
\ \ \ \ \ \ {\isacharparenleft}{\kern0pt}random{\isacharunderscore}{\kern0pt}seed{\isacharunderscore}{\kern0pt}xors\ {\isacharparenleft}{\kern0pt}find{\isacharunderscore}{\kern0pt}t\ {\isasymdelta}{\isacharparenright}{\kern0pt}\ {\isacharparenleft}{\kern0pt}length\ S{\isacharparenright}{\kern0pt}{\isacharparenright}{\kern0pt}{\isacharparenright}{\kern0pt}\isanewline
\ \ \ \ {\isacharbraceleft}{\kern0pt}c{\isachardot}{\kern0pt}\ projr\ c\ {\isasymin}\ {\isacharbraceleft}{\kern0pt}sz\ {\isacharslash}{\kern0pt}\ {\isacharparenleft}{\kern0pt}{\isadigit{1}}\ {\isacharplus}{\kern0pt}\ {\isasymepsilon}{\isacharparenright}{\kern0pt}\ {\isachardot}{\kern0pt}{\isachardot}{\kern0pt}\ {\isacharparenleft}{\kern0pt}{\isadigit{1}}\ {\isacharplus}{\kern0pt}\ {\isasymepsilon}{\isacharparenright}{\kern0pt}\ {\isacharasterisk}{\kern0pt}\ sz{\isacharbraceright}{\kern0pt}{\isacharbraceright}{\kern0pt}\ {\isasymge}\ {\isadigit{1}}\ {\isacharminus}{\kern0pt}\ {\isasymdelta}{\isachardoublequoteclose}%
}
\snip{appmclcertprob}{1}{2}{%
\isacommand{theorem}\isamarkupfalse%
\ certcheck{\isacharunderscore}{\kern0pt}prob{\isacharcolon}{\kern0pt}\isanewline
\isakeyword{assumes}\ {\isachardoublequoteopen}{\isacharparenleft}{\kern0pt}{\isasymAnd}F{\isachardot}{\kern0pt}\ check{\isacharunderscore}{\kern0pt}unsat\ F\ {\isasymLongrightarrow}\ sols\ F\ {\isacharequal}{\kern0pt}\ {\isacharbraceleft}{\kern0pt}{\isacharbraceright}{\kern0pt}{\isacharparenright}{\kern0pt}{\isachardoublequoteclose}\isanewline
\isakeyword{assumes}\ {\isachardoublequoteopen}{\isasymdelta}\ {\isachargreater}{\kern0pt}\ {\isadigit{0}}{\isachardoublequoteclose}\ {\isachardoublequoteopen}{\isasymdelta}\ {\isacharless}{\kern0pt}\ {\isadigit{1}}{\isachardoublequoteclose}\ {\isachardoublequoteopen}{\isasymepsilon}\ {\isachargreater}{\kern0pt}\ {\isadigit{0}}{\isachardoublequoteclose}\ {\isachardoublequoteopen}distinct\ S{\isachardoublequoteclose}\isanewline
\isakeyword{shows}\ {\isachardoublequoteopen}\isanewline
\ \ let\ sz\ {\isacharequal}{\kern0pt}\ real\ {\isacharparenleft}{\kern0pt}card\ {\isacharparenleft}{\kern0pt}proj\ {\isacharparenleft}{\kern0pt}set\ S{\isacharparenright}{\kern0pt}\ {\isacharparenleft}{\kern0pt}sols\ F{\isacharparenright}{\kern0pt}{\isacharparenright}{\kern0pt}{\isacharparenright}{\kern0pt}\ in\isanewline
\ \ let\ seeds\ {\isacharequal}{\kern0pt}\ random{\isacharunderscore}{\kern0pt}seed{\isacharunderscore}{\kern0pt}xors\ {\isacharparenleft}{\kern0pt}find{\isacharunderscore}{\kern0pt}t\ {\isasymdelta}{\isacharparenright}{\kern0pt}\ {\isacharparenleft}{\kern0pt}length\ S{\isacharparenright}{\kern0pt}\ in\ \isanewline
\ \ let\ pr\ {\isacharequal}{\kern0pt}\ measure{\isacharunderscore}{\kern0pt}pmf{\isachardot}{\kern0pt}prob\isanewline
\ \ \ \ {\isacharparenleft}{\kern0pt}map{\isacharunderscore}{\kern0pt}pmf\ {\isacharparenleft}{\kern0pt}{\isasymlambda}r{\isachardot}{\kern0pt}\ certcheck\ check{\isacharunderscore}{\kern0pt}unsat\ F\ S\ {\isasymepsilon}\ {\isasymdelta}\ {\isacharparenleft}{\kern0pt}f\ r{\isacharparenright}{\kern0pt}\ r{\isacharparenright}{\kern0pt}\ seeds{\isacharparenright}{\kern0pt}\ in\isanewline
\ \ pr\ {\isacharbraceleft}{\kern0pt}c{\isachardot}{\kern0pt}\ {\isasymnot}isl\ c\ {\isasymand}\ projr\ c\ {\isasymnotin}\ {\isacharbraceleft}{\kern0pt}sz\ {\isacharslash}{\kern0pt}\ {\isacharparenleft}{\kern0pt}{\isadigit{1}}\ {\isacharplus}{\kern0pt}\ {\isasymepsilon}{\isacharparenright}{\kern0pt}\ {\isachardot}{\kern0pt}{\isachardot}{\kern0pt}\ {\isacharparenleft}{\kern0pt}{\isadigit{1}}\ {\isacharplus}{\kern0pt}\ {\isasymepsilon}{\isacharparenright}{\kern0pt}\ {\isacharasterisk}{\kern0pt}\ sz{\isacharbraceright}{\kern0pt}{\isacharbraceright}{\kern0pt}\ {\isasymle}\ {\isasymdelta}\ {\isasymand}\isanewline
\ \ {\isacharparenleft}{\kern0pt}\ {\isacharparenleft}{\kern0pt}{\isasymforall}r\ {\isasymin}\ set{\isacharunderscore}{\kern0pt}pmf\ seeds{\isachardot}{\kern0pt}\isanewline
\ \ \ \ \ \ {\isasymnot}isl\ {\isacharparenleft}{\kern0pt}certcheck\ check{\isacharunderscore}{\kern0pt}unsat\ F\ S\ {\isasymepsilon}\ {\isasymdelta}\ {\isacharparenleft}{\kern0pt}f\ r{\isacharparenright}{\kern0pt}\ r{\isacharparenright}{\kern0pt}{\isacharparenright}{\kern0pt}\ {\isasymlongrightarrow}\isanewline
\ \ \ \ pr\ {\isacharbraceleft}{\kern0pt}c{\isachardot}{\kern0pt}\ projr\ c\ {\isasymin}\ {\isacharbraceleft}{\kern0pt}sz\ {\isacharslash}{\kern0pt}\ {\isacharparenleft}{\kern0pt}{\isadigit{1}}\ {\isacharplus}{\kern0pt}\ {\isasymepsilon}{\isacharparenright}{\kern0pt}\ {\isachardot}{\kern0pt}{\isachardot}{\kern0pt}\ {\isacharparenleft}{\kern0pt}{\isadigit{1}}\ {\isacharplus}{\kern0pt}\ {\isasymepsilon}{\isacharparenright}{\kern0pt}\ {\isacharasterisk}{\kern0pt}\ sz{\isacharbraceright}{\kern0pt}{\isacharbraceright}{\kern0pt}\ {\isasymge}\ {\isadigit{1}}\ {\isacharminus}{\kern0pt}\ {\isasymdelta}{\isacharparenright}{\kern0pt}{\isachardoublequoteclose}%
}
\snip{localeappmclcert}{1}{2}{%
\isacommand{locale}\isamarkupfalse%
\ CertCheck\ {\isacharequal}{\kern0pt}\ ApproxMCL\ sols\ enc{\isacharunderscore}{\kern0pt}xor\isanewline
\isakeyword{for}\ sols\ {\isacharcolon}{\kern0pt}{\isacharcolon}{\kern0pt}\ {\isachardoublequoteopen}{\isacharprime}{\kern0pt}fml\ {\isasymRightarrow}\ {\isacharparenleft}{\kern0pt}{\isacharprime}{\kern0pt}a\ {\isasymRightarrow}\ bool{\isacharparenright}{\kern0pt}\ set{\isachardoublequoteclose}\isanewline
\isakeyword{and}\ enc{\isacharunderscore}{\kern0pt}xor\ {\isacharcolon}{\kern0pt}{\isacharcolon}{\kern0pt}\ {\isachardoublequoteopen}{\isacharprime}{\kern0pt}a\ list\ {\isasymtimes}\ bool\ {\isasymRightarrow}\ {\isacharprime}{\kern0pt}fml\ {\isasymRightarrow}\ {\isacharprime}{\kern0pt}fml{\isachardoublequoteclose}\ {\isacharplus}{\kern0pt}\isanewline
\isakeyword{fixes}\ check{\isacharunderscore}{\kern0pt}sol\ {\isacharcolon}{\kern0pt}{\isacharcolon}{\kern0pt}\ {\isachardoublequoteopen}{\isacharprime}{\kern0pt}fml\ {\isasymRightarrow}\ {\isacharparenleft}{\kern0pt}{\isacharprime}{\kern0pt}a\ {\isasymRightarrow}\ bool{\isacharparenright}{\kern0pt}\ {\isasymRightarrow}\ bool{\isachardoublequoteclose}\isanewline
\isakeyword{fixes}\ ban{\isacharunderscore}{\kern0pt}sol\ {\isacharcolon}{\kern0pt}{\isacharcolon}{\kern0pt}\ {\isachardoublequoteopen}{\isacharprime}{\kern0pt}a\ sol\ {\isasymRightarrow}\ {\isacharprime}{\kern0pt}fml\ {\isasymRightarrow}\ {\isacharprime}{\kern0pt}fml{\isachardoublequoteclose}\isanewline
\isakeyword{assumes}\ {\isachardoublequoteopen}{\isasymAnd}F\ w{\isachardot}{\kern0pt}\ check{\isacharunderscore}{\kern0pt}sol\ F\ w\ {\isasymlongleftrightarrow}\ w\ {\isasymin}\ sols\ F{\isachardoublequoteclose}\isanewline
\isakeyword{assumes}\ {\isachardoublequoteopen}{\isasymAnd}F\ vs{\isachardot}{\kern0pt}\ sols\ {\isacharparenleft}{\kern0pt}ban{\isacharunderscore}{\kern0pt}sol\ vs\ F{\isacharparenright}{\kern0pt}\ {\isacharequal}{\kern0pt}\isanewline
\ \ sols\ F\ {\isasyminter}\ {\isacharbraceleft}{\kern0pt}{\isasymomega}{\isachardot}{\kern0pt}\ map\ {\isasymomega}\ {\isacharparenleft}{\kern0pt}map\ fst\ vs{\isacharparenright}{\kern0pt}\ {\isasymnoteq}\ map\ snd\ vs{\isacharbraceright}{\kern0pt}{\isachardoublequoteclose}%
}
\snip{paleyzigmund}{1}{2}{%
\isacommand{lemma}\isamarkupfalse%
\ paley{\isacharunderscore}{\kern0pt}zigmund{\isacharunderscore}{\kern0pt}inequality{\isacharcolon}{\kern0pt}\isanewline
\isakeyword{assumes}\ {\isachardoublequoteopen}finite\ {\isacharparenleft}{\kern0pt}set{\isacharunderscore}{\kern0pt}pmf\ p{\isacharparenright}{\kern0pt}{\isachardoublequoteclose}\ {\isachardoublequoteopen}{\isasymtheta}\ {\isasymle}\ {\isadigit{1}}{\isachardoublequoteclose}\isanewline
\isakeyword{assumes}\ {\isachardoublequoteopen}{\isasymAnd}y{\isachardot}{\kern0pt}\ y\ {\isasymin}\ set{\isacharunderscore}{\kern0pt}pmf\ p\ {\isasymLongrightarrow}\ Y\ y\ {\isasymge}\ {\isadigit{0}}{\isachardoublequoteclose}\isanewline
\isakeyword{shows}\ {\isachardoublequoteopen}\isanewline
\ \ measure{\isacharunderscore}{\kern0pt}pmf{\isachardot}{\kern0pt}prob\ p\ {\isacharbraceleft}{\kern0pt}y{\isachardot}{\kern0pt}\ Y\ y\ {\isachargreater}{\kern0pt}\ {\isasymtheta}\ {\isacharasterisk}{\kern0pt}\ measure{\isacharunderscore}{\kern0pt}pmf{\isachardot}{\kern0pt}expectation\ p\ Y{\isacharbraceright}{\kern0pt}\ {\isacharasterisk}{\kern0pt}\isanewline
\ \ {\isacharparenleft}{\kern0pt}measure{\isacharunderscore}{\kern0pt}pmf{\isachardot}{\kern0pt}variance\ p\ Y\ {\isacharplus}{\kern0pt}\ {\isacharparenleft}{\kern0pt}{\isadigit{1}}{\isacharminus}{\kern0pt}{\isasymtheta}{\isacharparenright}{\kern0pt}{\isacharcircum}{\kern0pt}{\isadigit{2}}\ {\isacharasterisk}{\kern0pt}\ {\isacharparenleft}{\kern0pt}measure{\isacharunderscore}{\kern0pt}pmf{\isachardot}{\kern0pt}expectation\ p\ Y{\isacharparenright}{\kern0pt}{\isacharcircum}{\kern0pt}{\isadigit{2}}{\isacharparenright}{\kern0pt}\isanewline
\ \ \ \ {\isasymge}\ {\isacharparenleft}{\kern0pt}{\isadigit{1}}{\isacharminus}{\kern0pt}{\isasymtheta}{\isacharparenright}{\kern0pt}{\isacharcircum}{\kern0pt}{\isadigit{2}}\ {\isacharasterisk}{\kern0pt}\ {\isacharparenleft}{\kern0pt}measure{\isacharunderscore}{\kern0pt}pmf{\isachardot}{\kern0pt}expectation\ p\ Y{\isacharparenright}{\kern0pt}{\isacharcircum}{\kern0pt}{\isadigit{2}}{\isachardoublequoteclose}%
}
\snip{xorhashthreeuniv}{1}{2}{%
\isacommand{theorem}\isamarkupfalse%
\ xor{\isacharunderscore}{\kern0pt}hash{\isacharunderscore}{\kern0pt}family{\isacharunderscore}{\kern0pt}{\isadigit{3}}{\isacharunderscore}{\kern0pt}universal{\isacharcolon}{\kern0pt}\isanewline
\isakeyword{assumes}\ {\isachardoublequoteopen}finite\ V{\isachardoublequoteclose}\isanewline
\isakeyword{shows}{\isachardoublequoteopen}\isanewline
\ \ prob{\isacharunderscore}{\kern0pt}space{\isachardot}{\kern0pt}k{\isacharunderscore}{\kern0pt}universal\ {\isacharparenleft}{\kern0pt}random{\isacharunderscore}{\kern0pt}xors\ V\ n{\isacharparenright}{\kern0pt}\ {\isadigit{3}}\ xor{\isacharunderscore}{\kern0pt}hash\isanewline
\ \ {\isacharbraceleft}{\kern0pt}{\isasymalpha}{\isachardot}{\kern0pt}\ dom\ {\isasymalpha}\ {\isacharequal}{\kern0pt}\ V{\isacharbraceright}{\kern0pt}\ {\isacharbraceleft}{\kern0pt}{\isasymalpha}{\isachardot}{\kern0pt}\ dom\ {\isasymalpha}\ {\isacharequal}{\kern0pt}\ {\isacharbraceleft}{\kern0pt}{\isadigit{0}}{\isachardot}{\kern0pt}{\isachardot}{\kern0pt}{\isacharless}{\kern0pt}n{\isacharbraceright}{\kern0pt}{\isacharbraceright}{\kern0pt}{\isachardoublequoteclose}%
}
\snip{xorhashfamily}{1}{2}{%
\isacommand{definition}\isamarkupfalse%
\ xor{\isacharunderscore}{\kern0pt}hash\ \isakeyword{where}\ {\isachardoublequoteopen}\isanewline
\ \ xor{\isacharunderscore}{\kern0pt}hash\ w\ xors\ {\isacharequal}{\kern0pt}\isanewline
\ \ {\isacharparenleft}{\kern0pt}map{\isacharunderscore}{\kern0pt}option\ {\isacharparenleft}{\kern0pt}{\isasymlambda}xor{\isachardot}{\kern0pt}\ satisfies{\isacharunderscore}{\kern0pt}xor\ xor\ {\isacharbraceleft}{\kern0pt}x{\isachardot}{\kern0pt}\ w\ x\ {\isacharequal}{\kern0pt}\ Some\ True{\isacharbraceright}{\kern0pt}{\isacharparenright}{\kern0pt}\ {\isasymcirc}\ xors{\isacharparenright}{\kern0pt}{\isachardoublequoteclose}%
}
\snip{randomxor}{1}{2}{%
\isacommand{definition}\isamarkupfalse%
\ random{\isacharunderscore}{\kern0pt}xor\ \isakeyword{where}\ {\isachardoublequoteopen}\isanewline
\ \ random{\isacharunderscore}{\kern0pt}xor\ V\ {\isacharequal}{\kern0pt}\ pair{\isacharunderscore}{\kern0pt}pmf\ {\isacharparenleft}{\kern0pt}pmf{\isacharunderscore}{\kern0pt}of{\isacharunderscore}{\kern0pt}set\ {\isacharparenleft}{\kern0pt}Pow\ V{\isacharparenright}{\kern0pt}{\isacharparenright}{\kern0pt}\ {\isacharparenleft}{\kern0pt}bernoulli{\isacharunderscore}{\kern0pt}pmf\ {\isacharparenleft}{\kern0pt}{\isadigit{1}}{\isacharslash}{\kern0pt}{\isadigit{2}}{\isacharparenright}{\kern0pt}{\isacharparenright}{\kern0pt}{\isachardoublequoteclose}\isanewline
\isacommand{definition}\isamarkupfalse%
\ random{\isacharunderscore}{\kern0pt}xors\ \isakeyword{where}\ {\isachardoublequoteopen}\isanewline
\ \ random{\isacharunderscore}{\kern0pt}xors\ V\ n\ {\isacharequal}{\kern0pt}\ prod{\isacharunderscore}{\kern0pt}pmf\ {\isacharbraceleft}{\kern0pt}{\isachardot}{\kern0pt}{\isachardot}{\kern0pt}{\isacharless}{\kern0pt}n{\isacharbraceright}{\kern0pt}\ {\isacharparenleft}{\kern0pt}{\isasymlambda}{\isacharunderscore}{\kern0pt}{\isachardot}{\kern0pt}\ map{\isacharunderscore}{\kern0pt}pmf\ Some\ {\isacharparenleft}{\kern0pt}random{\isacharunderscore}{\kern0pt}xor\ V{\isacharparenright}{\kern0pt}{\isacharparenright}{\kern0pt}{\isachardoublequoteclose}%
}
\snip{approxmcdef}{1}{2}{%
\isacommand{definition}\isamarkupfalse%
\ approxmccore\ {\isacharcolon}{\kern0pt}{\isacharcolon}{\kern0pt}\ {\isachardoublequoteopen}{\isacharprime}{\kern0pt}fml\ {\isasymRightarrow}\ {\isacharprime}{\kern0pt}a\ set\ {\isasymRightarrow}\ nat\ {\isasymRightarrow}\ nat\ pmf{\isachardoublequoteclose}\isanewline
\isakeyword{where}\ {\isachardoublequoteopen}approxmccore\ F\ S\ thresh\ {\isacharequal}{\kern0pt}\isanewline
\ \ map{\isacharunderscore}{\kern0pt}pmf\ {\isacharparenleft}{\kern0pt}approxcore{\isacharunderscore}{\kern0pt}xors\ F\ S\ thresh{\isacharparenright}{\kern0pt}\ {\isacharparenleft}{\kern0pt}random{\isacharunderscore}{\kern0pt}xors\ S\ {\isacharparenleft}{\kern0pt}card\ S\ {\isacharminus}{\kern0pt}\ {\isadigit{1}}{\isacharparenright}{\kern0pt}{\isacharparenright}{\kern0pt}{\isachardoublequoteclose}\isanewline
\isanewline
\isacommand{definition}\isamarkupfalse%
\ approxmc{\isacharcolon}{\kern0pt}{\isacharcolon}{\kern0pt}{\isachardoublequoteopen}{\isacharprime}{\kern0pt}fml\ {\isasymRightarrow}\ {\isacharprime}{\kern0pt}a\ set\ {\isasymRightarrow}\ real\ {\isasymRightarrow}\ real\ {\isasymRightarrow}\ nat\ {\isasymRightarrow}\ nat\ pmf{\isachardoublequoteclose}\isanewline
\isakeyword{where}\ {\isachardoublequoteopen}approxmc\ F\ S\ {\isasymepsilon}\ {\isasymdelta}\ n\ {\isacharequal}{\kern0pt}\ {\isacharparenleft}{\kern0pt}\isanewline
\ \ let\ thresh\ {\isacharequal}{\kern0pt}\ compute{\isacharunderscore}{\kern0pt}thresh\ {\isasymepsilon}\ in\isanewline
\ \ if\ card\ {\isacharparenleft}{\kern0pt}proj\ S\ {\isacharparenleft}{\kern0pt}sols\ F{\isacharparenright}{\kern0pt}{\isacharparenright}{\kern0pt}\ {\isacharless}{\kern0pt}\ thresh\isanewline
\ \ then\ return{\isacharunderscore}{\kern0pt}pmf\ {\isacharparenleft}{\kern0pt}card\ {\isacharparenleft}{\kern0pt}proj\ S\ {\isacharparenleft}{\kern0pt}sols\ F{\isacharparenright}{\kern0pt}{\isacharparenright}{\kern0pt}{\isacharparenright}{\kern0pt}\isanewline
\ \ else\isanewline
\ \ \ \ let\ t\ {\isacharequal}{\kern0pt}\ compute{\isacharunderscore}{\kern0pt}t\ {\isasymdelta}\ n\ in\isanewline
\ \ \ \ map{\isacharunderscore}{\kern0pt}pmf\ {\isacharparenleft}{\kern0pt}median\ t{\isacharparenright}{\kern0pt}\isanewline
\ \ \ \ \ \ {\isacharparenleft}{\kern0pt}prod{\isacharunderscore}{\kern0pt}pmf\ {\isacharbraceleft}{\kern0pt}{\isadigit{0}}{\isachardot}{\kern0pt}{\isachardot}{\kern0pt}{\isacharless}{\kern0pt}t{\isacharcolon}{\kern0pt}{\isacharcolon}{\kern0pt}nat{\isacharbraceright}{\kern0pt}\ {\isacharparenleft}{\kern0pt}{\isasymlambda}i{\isachardot}{\kern0pt}\ approxmccore\ F\ S\ thresh{\isacharparenright}{\kern0pt}{\isacharparenright}{\kern0pt}{\isacharparenright}{\kern0pt}{\isachardoublequoteclose}%
}
\snip{TLUdef}{1}{2}{%
\isacommand{definition}\isamarkupfalse%
\ {\isasymmu}\ \isakeyword{where}\ {\isachardoublequoteopen}{\isasymmu}\ i\ {\isacharequal}{\kern0pt}\ card\ {\isacharparenleft}{\kern0pt}proj\ S\ W{\isacharparenright}{\kern0pt}\ {\isacharslash}{\kern0pt}\ {\isadigit{2}}\ {\isacharcircum}{\kern0pt}\ i{\isachardoublequoteclose}\isanewline
\isacommand{definition}\isamarkupfalse%
\ T\ \isakeyword{where}\ {\isachardoublequoteopen}T\ i\ {\isacharequal}{\kern0pt}\ {\isacharbraceleft}{\kern0pt}h{\isachardot}{\kern0pt}\ card{\isacharunderscore}{\kern0pt}slice\ h\ i\ {\isacharless}{\kern0pt}\ thresh{\isacharbraceright}{\kern0pt}{\isachardoublequoteclose}\isanewline
\isacommand{definition}\isamarkupfalse%
\ L\ \isakeyword{where}\ {\isachardoublequoteopen}L\ i\ {\isacharequal}{\kern0pt}\ {\isacharbraceleft}{\kern0pt}h{\isachardot}{\kern0pt}\ card{\isacharunderscore}{\kern0pt}slice\ h\ i\ {\isacharless}{\kern0pt}\ {\isasymmu}\ i\ {\isacharslash}{\kern0pt}\ {\isacharparenleft}{\kern0pt}{\isadigit{1}}{\isacharplus}{\kern0pt}{\isasymepsilon}{\isacharparenright}{\kern0pt}{\isacharbraceright}{\kern0pt}{\isachardoublequoteclose}\isanewline
\isacommand{definition}\isamarkupfalse%
\ U\ \isakeyword{where}\ {\isachardoublequoteopen}U\ i\ {\isacharequal}{\kern0pt}\ {\isacharbraceleft}{\kern0pt}h{\isachardot}{\kern0pt}\ card{\isacharunderscore}{\kern0pt}slice\ h\ i\ {\isasymge}\ {\isasymmu}\ i\ {\isacharasterisk}{\kern0pt}\ {\isacharparenleft}{\kern0pt}{\isadigit{1}}\ {\isacharplus}{\kern0pt}\ {\isasymepsilon}{\isacharslash}{\kern0pt}{\isacharparenleft}{\kern0pt}{\isadigit{1}}{\isacharplus}{\kern0pt}{\isasymepsilon}{\isacharparenright}{\kern0pt}{\isacharparenright}{\kern0pt}{\isacharbraceright}{\kern0pt}{\isachardoublequoteclose}%
}
\snip{appcore}{1}{2}{%
\isacommand{definition}\isamarkupfalse%
\ approxcore\ \isakeyword{where}\ {\isachardoublequoteopen}\isanewline
\ \ approxcore\ h\ {\isacharequal}{\kern0pt}\ {\isacharparenleft}{\kern0pt}\isanewline
\ \ case\ List{\isachardot}{\kern0pt}find\ {\isacharparenleft}{\kern0pt}{\isasymlambda}i{\isachardot}{\kern0pt}\ h\ {\isasymin}\ T\ i{\isacharparenright}{\kern0pt}\ {\isacharbrackleft}{\kern0pt}{\isadigit{1}}{\isachardot}{\kern0pt}{\isachardot}{\kern0pt}{\isacharless}{\kern0pt}card\ S{\isacharbrackright}{\kern0pt}\ of\isanewline
\ \ \ \ None\ {\isasymRightarrow}\ {\isacharparenleft}{\kern0pt}{\isadigit{2}}\ {\isacharcircum}{\kern0pt}\ card\ S{\isacharcomma}{\kern0pt}\ {\isadigit{1}}{\isacharparenright}{\kern0pt}\isanewline
\ \ {\isacharbar}{\kern0pt}\ Some\ m\ {\isasymRightarrow}\ {\isacharparenleft}{\kern0pt}{\isadigit{2}}\ {\isacharcircum}{\kern0pt}\ m{\isacharcomma}{\kern0pt}\ card{\isacharunderscore}{\kern0pt}slice\ h\ m{\isacharparenright}{\kern0pt}{\isacharparenright}{\kern0pt}{\isachardoublequoteclose}\isanewline
\isanewline
\isacommand{definition}\isamarkupfalse%
\ approxcore{\isacharunderscore}{\kern0pt}fail\ \isakeyword{where}\ {\isachardoublequoteopen}\isanewline
\ \ approxcore{\isacharunderscore}{\kern0pt}fail\ \ {\isacharequal}{\kern0pt}\isanewline
\ \ {\isacharbraceleft}{\kern0pt}h{\isachardot}{\kern0pt}\ let\ {\isacharparenleft}{\kern0pt}cells{\isacharcomma}{\kern0pt}sols{\isacharparenright}{\kern0pt}\ {\isacharequal}{\kern0pt}\ approxcore\ h\ {\isacharsemicolon}{\kern0pt}\ sz\ {\isacharequal}{\kern0pt}\ card\ {\isacharparenleft}{\kern0pt}proj\ S\ W{\isacharparenright}{\kern0pt}\ in\isanewline
\ \ \ \ cells\ {\isacharasterisk}{\kern0pt}\ sols\ {\isasymnotin}\ {\isacharbraceleft}{\kern0pt}sz\ {\isacharslash}{\kern0pt}\ {\isacharparenleft}{\kern0pt}{\isadigit{1}}\ {\isacharplus}{\kern0pt}\ {\isasymepsilon}{\isacharparenright}{\kern0pt}\ {\isachardot}{\kern0pt}{\isachardot}{\kern0pt}\ {\isacharparenleft}{\kern0pt}{\isadigit{1}}\ {\isacharplus}{\kern0pt}\ {\isasymepsilon}{\isacharparenright}{\kern0pt}\ {\isacharasterisk}{\kern0pt}\ sz{\isacharbraceright}{\kern0pt}{\isacharbraceright}{\kern0pt}{\isachardoublequoteclose}%
}
\snip{appcorefailprob}{1}{2}{%
\isacommand{lemma}\isamarkupfalse%
\ approxcore{\isacharunderscore}{\kern0pt}fail{\isacharunderscore}{\kern0pt}prob{\isacharcolon}{\kern0pt}\isanewline
\isakeyword{assumes}\ {\isachardoublequoteopen}{\isacharparenleft}{\kern0pt}{\isadigit{1}}\ {\isacharplus}{\kern0pt}\ {\isasymepsilon}\ {\isacharslash}{\kern0pt}\ {\isacharparenleft}{\kern0pt}{\isadigit{1}}\ {\isacharplus}{\kern0pt}\ {\isasymepsilon}{\isacharparenright}{\kern0pt}{\isacharparenright}{\kern0pt}\ {\isacharasterisk}{\kern0pt}\ {\isacharparenleft}{\kern0pt}{\isadigit{9}}{\isachardot}{\kern0pt}{\isadigit{8}}{\isadigit{4}}\ {\isacharasterisk}{\kern0pt}\ {\isacharparenleft}{\kern0pt}{\isadigit{1}}\ {\isacharplus}{\kern0pt}\ {\isadigit{1}}\ {\isacharslash}{\kern0pt}\ {\isasymepsilon}{\isacharparenright}{\kern0pt}{\isacharcircum}{\kern0pt}{\isadigit{2}}{\isacharparenright}{\kern0pt}\ {\isasymle}\ thresh{\isachardoublequoteclose}\isanewline
\isakeyword{assumes}\ {\isachardoublequoteopen}{\isasymepsilon}\ {\isasymle}\ {\isadigit{1}}{\isachardoublequoteclose}\ {\isachardoublequoteopen}finite\ {\isacharparenleft}{\kern0pt}set{\isacharunderscore}{\kern0pt}pmf\ p{\isacharparenright}{\kern0pt}{\isachardoublequoteclose}\isanewline
\isakeyword{assumes}\ {\isachardoublequoteopen}prob{\isacharunderscore}{\kern0pt}space{\isachardot}{\kern0pt}k{\isacharunderscore}{\kern0pt}universal\ {\isacharparenleft}{\kern0pt}measure{\isacharunderscore}{\kern0pt}pmf\ p{\isacharparenright}{\kern0pt}\ {\isadigit{2}}\ H\isanewline
\ \ \ \ \ \ \ \ \ \ {\isacharbraceleft}{\kern0pt}{\isasymalpha}{\isachardot}{\kern0pt}\ dom\ {\isasymalpha}\ {\isacharequal}{\kern0pt}\ S{\isacharbraceright}{\kern0pt}\ {\isacharbraceleft}{\kern0pt}{\isasymalpha}{\isachardot}{\kern0pt}\ dom\ {\isasymalpha}\ {\isacharequal}{\kern0pt}\ {\isacharbraceleft}{\kern0pt}{\isadigit{0}}{\isachardot}{\kern0pt}{\isachardot}{\kern0pt}{\isacharless}{\kern0pt}card\ S\ {\isacharminus}{\kern0pt}\ {\isadigit{1}}{\isacharbraceright}{\kern0pt}{\isacharbraceright}{\kern0pt}{\isachardoublequoteclose}\isanewline
\isakeyword{shows}\ {\isachardoublequoteopen}\isanewline
\ \ measure{\isacharunderscore}{\kern0pt}pmf{\isachardot}{\kern0pt}prob\ {\isacharparenleft}{\kern0pt}map{\isacharunderscore}{\kern0pt}pmf\ {\isacharparenleft}{\kern0pt}{\isasymlambda}s\ w{\isachardot}{\kern0pt}\ H\ w\ s{\isacharparenright}{\kern0pt}\ p{\isacharparenright}{\kern0pt}\ approxcore{\isacharunderscore}{\kern0pt}fail\ {\isasymle}\ {\isadigit{0}}{\isachardot}{\kern0pt}{\isadigit{3}}{\isadigit{6}}{\isachardoublequoteclose}%
}
\snip{mstarexists}{1}{2}{%
\isacommand{lemma}\isamarkupfalse%
\ mstar{\isacharunderscore}{\kern0pt}exists{\isacharcolon}{\kern0pt}\isanewline
\isakeyword{obtains}\ mstar\ \isakeyword{where}\isanewline
\ \ {\isachardoublequoteopen}{\isasymmu}\ {\isacharparenleft}{\kern0pt}mstar\ {\isacharminus}{\kern0pt}\ {\isadigit{1}}{\isacharparenright}{\kern0pt}\ {\isacharasterisk}{\kern0pt}\ {\isacharparenleft}{\kern0pt}{\isadigit{1}}\ {\isacharplus}{\kern0pt}\ {\isasymepsilon}\ {\isacharslash}{\kern0pt}\ {\isacharparenleft}{\kern0pt}{\isadigit{1}}\ {\isacharplus}{\kern0pt}\ {\isasymepsilon}{\isacharparenright}{\kern0pt}{\isacharparenright}{\kern0pt}\ {\isachargreater}{\kern0pt}\ thresh{\isachardoublequoteclose}\isanewline
\ \ {\isachardoublequoteopen}{\isasymmu}\ mstar\ {\isacharasterisk}{\kern0pt}\ {\isacharparenleft}{\kern0pt}{\isadigit{1}}\ {\isacharplus}{\kern0pt}\ {\isasymepsilon}\ {\isacharslash}{\kern0pt}\ {\isacharparenleft}{\kern0pt}{\isadigit{1}}\ {\isacharplus}{\kern0pt}\ {\isasymepsilon}{\isacharparenright}{\kern0pt}{\isacharparenright}{\kern0pt}\ {\isasymle}\ thresh{\isachardoublequoteclose}\isanewline
\ \ {\isachardoublequoteopen}mstar\ {\isasymle}\ card\ S\ {\isacharminus}{\kern0pt}\ {\isadigit{1}}{\isachardoublequoteclose}%
}
\snip{failuresubset}{1}{2}{%
\isacommand{lemma}\isamarkupfalse%
\ failure{\isacharunderscore}{\kern0pt}subset{\isacharcolon}{\kern0pt}\ \ \ \isanewline
\isakeyword{shows}\ {\isachardoublequoteopen}approxcore{\isacharunderscore}{\kern0pt}fail\ {\isasymsubseteq}\isanewline
\ \ T\ {\isacharparenleft}{\kern0pt}mstar{\isacharminus}{\kern0pt}{\isadigit{3}}{\isacharparenright}{\kern0pt}\ {\isasymunion}\ L\ {\isacharparenleft}{\kern0pt}mstar{\isacharminus}{\kern0pt}{\isadigit{2}}{\isacharparenright}{\kern0pt}\ {\isasymunion}\ L\ {\isacharparenleft}{\kern0pt}mstar{\isacharminus}{\kern0pt}{\isadigit{1}}{\isacharparenright}{\kern0pt}\ {\isasymunion}\ {\isacharparenleft}{\kern0pt}L\ mstar\ {\isasymunion}\ U\ mstar{\isacharparenright}{\kern0pt}{\isachardoublequoteclose}%
}
\snip{eventsprob}{1}{2}{%
\isacommand{lemma}\isamarkupfalse%
\ events{\isacharunderscore}{\kern0pt}prob{\isacharcolon}{\kern0pt}\isanewline
\isakeyword{assumes}\ {\isachardoublequoteopen}{\isacharparenleft}{\kern0pt}{\isadigit{1}}\ {\isacharplus}{\kern0pt}\ {\isasymepsilon}\ {\isacharslash}{\kern0pt}\ {\isacharparenleft}{\kern0pt}{\isadigit{1}}\ {\isacharplus}{\kern0pt}\ {\isasymepsilon}{\isacharparenright}{\kern0pt}{\isacharparenright}{\kern0pt}\ {\isacharasterisk}{\kern0pt}\ {\isacharparenleft}{\kern0pt}{\isadigit{9}}{\isachardot}{\kern0pt}{\isadigit{8}}{\isadigit{4}}\ {\isacharasterisk}{\kern0pt}\ {\isacharparenleft}{\kern0pt}{\isadigit{1}}\ {\isacharplus}{\kern0pt}\ {\isadigit{1}}\ {\isacharslash}{\kern0pt}\ {\isasymepsilon}{\isacharparenright}{\kern0pt}{\isacharcircum}{\kern0pt}{\isadigit{2}}{\isacharparenright}{\kern0pt}\ {\isasymle}\ thresh{\isachardoublequoteclose}\isanewline
\isakeyword{assumes}\ {\isachardoublequoteopen}finite\ {\isacharparenleft}{\kern0pt}set{\isacharunderscore}{\kern0pt}pmf\ p{\isacharparenright}{\kern0pt}{\isachardoublequoteclose}\isanewline
\isakeyword{assumes}\ {\isachardoublequoteopen}prob{\isacharunderscore}{\kern0pt}space{\isachardot}{\kern0pt}k{\isacharunderscore}{\kern0pt}universal\ {\isacharparenleft}{\kern0pt}measure{\isacharunderscore}{\kern0pt}pmf\ p{\isacharparenright}{\kern0pt}\ {\isadigit{2}}\ H\isanewline
\ \ \ \ \ \ \ \ \ \ {\isacharbraceleft}{\kern0pt}{\isasymalpha}{\isachardot}{\kern0pt}\ dom\ {\isasymalpha}\ {\isacharequal}{\kern0pt}\ S{\isacharbraceright}{\kern0pt}\ {\isacharbraceleft}{\kern0pt}{\isasymalpha}{\isachardot}{\kern0pt}\ dom\ {\isasymalpha}\ {\isacharequal}{\kern0pt}\ {\isacharbraceleft}{\kern0pt}{\isadigit{0}}{\isachardot}{\kern0pt}{\isachardot}{\kern0pt}{\isacharless}{\kern0pt}card\ S\ {\isacharminus}{\kern0pt}\ {\isadigit{1}}{\isacharbraceright}{\kern0pt}{\isacharbraceright}{\kern0pt}{\isachardoublequoteclose}\isanewline
\isakeyword{shows}\ {\isachardoublequoteopen}let\ Hp\ {\isacharequal}{\kern0pt}\ map{\isacharunderscore}{\kern0pt}pmf\ {\isacharparenleft}{\kern0pt}{\isasymlambda}s\ w{\isachardot}{\kern0pt}\ H\ w\ s{\isacharparenright}{\kern0pt}\ p\ in\isanewline
\ \ {\isacharparenleft}{\kern0pt}{\isasymepsilon}\ {\isasymle}\ {\isadigit{1}}\ {\isasymlongrightarrow}\ measure{\isacharunderscore}{\kern0pt}pmf{\isachardot}{\kern0pt}prob\ Hp\ {\isacharparenleft}{\kern0pt}T\ {\isacharparenleft}{\kern0pt}mstar{\isacharminus}{\kern0pt}{\isadigit{3}}{\isacharparenright}{\kern0pt}{\isacharparenright}{\kern0pt}\ {\isasymle}\ {\isadigit{1}}\ {\isacharslash}{\kern0pt}\ {\isadigit{6}}{\isadigit{2}}{\isachardot}{\kern0pt}{\isadigit{5}}{\isacharparenright}{\kern0pt}\ {\isasymand}\isanewline
\ \ measure{\isacharunderscore}{\kern0pt}pmf{\isachardot}{\kern0pt}prob\ Hp\ {\isacharparenleft}{\kern0pt}L\ {\isacharparenleft}{\kern0pt}mstar{\isacharminus}{\kern0pt}{\isadigit{2}}{\isacharparenright}{\kern0pt}{\isacharparenright}{\kern0pt}\ {\isasymle}\ {\isadigit{1}}\ {\isacharslash}{\kern0pt}\ {\isadigit{2}}{\isadigit{0}}{\isachardot}{\kern0pt}{\isadigit{6}}{\isadigit{8}}\ {\isasymand}\isanewline
\ \ measure{\isacharunderscore}{\kern0pt}pmf{\isachardot}{\kern0pt}prob\ Hp\ {\isacharparenleft}{\kern0pt}L\ {\isacharparenleft}{\kern0pt}mstar{\isacharminus}{\kern0pt}{\isadigit{1}}{\isacharparenright}{\kern0pt}{\isacharparenright}{\kern0pt}\ {\isasymle}\ {\isadigit{1}}\ {\isacharslash}{\kern0pt}\ {\isadigit{1}}{\isadigit{0}}{\isachardot}{\kern0pt}{\isadigit{8}}{\isadigit{4}}\ {\isasymand}\isanewline
\ \ measure{\isacharunderscore}{\kern0pt}pmf{\isachardot}{\kern0pt}prob\ Hp\ {\isacharparenleft}{\kern0pt}L\ mstar\ {\isasymunion}\ U\ mstar{\isacharparenright}{\kern0pt}\ {\isasymle}\ {\isadigit{1}}\ {\isacharslash}{\kern0pt}\ {\isadigit{4}}{\isachardot}{\kern0pt}{\isadigit{9}}{\isadigit{2}}{\isachardoublequoteclose}%
}
\snip{appmcprob}{1}{2}{%
\isacommand{theorem}\isamarkupfalse%
\ approxmc{\isacharunderscore}{\kern0pt}prob{\isacharcolon}{\kern0pt}\isanewline
\isakeyword{assumes}\ {\isachardoublequoteopen}{\isasymdelta}\ {\isachargreater}{\kern0pt}\ {\isadigit{0}}{\isachardoublequoteclose}\ {\isachardoublequoteopen}{\isasymdelta}\ {\isacharless}{\kern0pt}\ {\isadigit{1}}{\isachardoublequoteclose}\ {\isachardoublequoteopen}{\isasymepsilon}\ {\isachargreater}{\kern0pt}\ {\isadigit{0}}{\isachardoublequoteclose}\ {\isachardoublequoteopen}{\isasymepsilon}\ {\isasymle}\ {\isadigit{1}}{\isachardoublequoteclose}\ {\isachardoublequoteopen}finite\ S{\isachardoublequoteclose}\isanewline
\isakeyword{shows}\ {\isachardoublequoteopen}let\ sz\ {\isacharequal}{\kern0pt}\ real\ {\isacharparenleft}{\kern0pt}card\ {\isacharparenleft}{\kern0pt}proj\ S\ {\isacharparenleft}{\kern0pt}sols\ F{\isacharparenright}{\kern0pt}{\isacharparenright}{\kern0pt}{\isacharparenright}{\kern0pt}\ in\isanewline
\ \ measure{\isacharunderscore}{\kern0pt}pmf{\isachardot}{\kern0pt}prob\ {\isacharparenleft}{\kern0pt}approxmc\ F\ S\ {\isasymepsilon}\ {\isasymdelta}\ n{\isacharparenright}{\kern0pt}\isanewline
\ \ \ \ {\isacharbraceleft}{\kern0pt}c{\isachardot}{\kern0pt}\ c\ {\isasymin}\ {\isacharbraceleft}{\kern0pt}sz\ {\isacharslash}{\kern0pt}\ {\isacharparenleft}{\kern0pt}{\isadigit{1}}\ {\isacharplus}{\kern0pt}\ {\isasymepsilon}{\isacharparenright}{\kern0pt}\ {\isachardot}{\kern0pt}{\isachardot}{\kern0pt}\ {\isacharparenleft}{\kern0pt}{\isadigit{1}}\ {\isacharplus}{\kern0pt}\ {\isasymepsilon}{\isacharparenright}{\kern0pt}\ {\isacharasterisk}{\kern0pt}\ sz{\isacharbraceright}{\kern0pt}{\isacharbraceright}{\kern0pt}\ {\isasymge}\ {\isadigit{1}}\ {\isacharminus}{\kern0pt}\ {\isasymdelta}{\isachardoublequoteclose}%
}
\snip{localeappmc}{1}{2}{%
\isacommand{locale}\isamarkupfalse%
\ ApproxMC\ {\isacharequal}{\kern0pt}\isanewline
\isakeyword{fixes}\ sols\ {\isacharcolon}{\kern0pt}{\isacharcolon}{\kern0pt}\ {\isachardoublequoteopen}{\isacharprime}{\kern0pt}fml\ {\isasymRightarrow}\ {\isacharparenleft}{\kern0pt}{\isacharprime}{\kern0pt}a\ {\isasymRightarrow}\ bool{\isacharparenright}{\kern0pt}\ set{\isachardoublequoteclose}\isanewline
\isakeyword{fixes}\ enc{\isacharunderscore}{\kern0pt}xor\ {\isacharcolon}{\kern0pt}{\isacharcolon}{\kern0pt}\ {\isachardoublequoteopen}{\isacharprime}{\kern0pt}a\ set\ {\isasymtimes}\ bool\ {\isasymRightarrow}\ {\isacharprime}{\kern0pt}fml\ {\isasymRightarrow}\ {\isacharprime}{\kern0pt}fml{\isachardoublequoteclose}\isanewline
\isakeyword{assumes}\ {\isachardoublequoteopen}{\isasymAnd}F\ xor{\isachardot}{\kern0pt}\isanewline
\ \ sols\ {\isacharparenleft}{\kern0pt}enc{\isacharunderscore}{\kern0pt}xor\ xor\ F{\isacharparenright}{\kern0pt}\ {\isacharequal}{\kern0pt}\ sols\ F\ {\isasyminter}\ {\isacharbraceleft}{\kern0pt}{\isasymomega}{\isachardot}{\kern0pt}\ satisfies{\isacharunderscore}{\kern0pt}xor\ xor\ {\isacharbraceleft}{\kern0pt}x{\isachardot}{\kern0pt}\ {\isasymomega}\ x{\isacharbraceright}{\kern0pt}{\isacharbraceright}{\kern0pt}{\isachardoublequoteclose}%
}
\snip{chebyshev}{1}{2}{%
\isacommand{lemma}\isamarkupfalse%
\ chebyshev{\isacharunderscore}{\kern0pt}inequality{\isacharcolon}{\kern0pt}\isanewline
\isakeyword{assumes}\ {\isachardoublequoteopen}k\ {\isachargreater}{\kern0pt}\ {\isadigit{0}}{\isachardoublequoteclose}\ {\isachardoublequoteopen}measure{\isacharunderscore}{\kern0pt}pmf{\isachardot}{\kern0pt}variance\ p\ Y\ {\isachargreater}{\kern0pt}\ {\isadigit{0}}{\isachardoublequoteclose}\isanewline
\isakeyword{assumes}\ {\isachardoublequoteopen}integrable\ p\ {\isacharparenleft}{\kern0pt}{\isasymlambda}y{\isachardot}{\kern0pt}\ {\isacharparenleft}{\kern0pt}Y\ y{\isacharparenright}{\kern0pt}\isactrlsup {\isadigit{2}}{\isacharparenright}{\kern0pt}{\isachardoublequoteclose}\isanewline
\isakeyword{shows}\ {\isachardoublequoteopen}\isanewline
\ \ measure{\isacharunderscore}{\kern0pt}pmf{\isachardot}{\kern0pt}prob\ p\isanewline
\ \ {\isacharbraceleft}{\kern0pt}y{\isachardot}{\kern0pt}\ {\isacharparenleft}{\kern0pt}Y\ y\ {\isacharminus}{\kern0pt}\ measure{\isacharunderscore}{\kern0pt}pmf{\isachardot}{\kern0pt}expectation\ p\ Y{\isacharparenright}{\kern0pt}\isactrlsup {\isadigit{2}}\isanewline
\ \ \ \ \ \ \ \ {\isasymge}\ k\isactrlsup {\isadigit{2}}\ {\isacharasterisk}{\kern0pt}\ measure{\isacharunderscore}{\kern0pt}pmf{\isachardot}{\kern0pt}variance\ p\ Y{\isacharbraceright}{\kern0pt}\ {\isasymle}\ {\isadigit{1}}\ {\isacharslash}{\kern0pt}\ k\isactrlsup {\isadigit{2}}{\isachardoublequoteclose}%
}
\snip{localemeasurepmf}{1}{2}{%
\isacommand{interpretation}\isamarkupfalse%
\ measure{\isacharunderscore}{\kern0pt}pmf{\isacharcolon}{\kern0pt}\ prob{\isacharunderscore}{\kern0pt}space\ {\isachardoublequoteopen}measure{\isacharunderscore}{\kern0pt}pmf\ p{\isachardoublequoteclose}%
}
\snip{localeprobspace}{1}{2}{%
\isacommand{locale}\isamarkupfalse%
\ prob{\isacharunderscore}{\kern0pt}space\ {\isacharequal}{\kern0pt}\ finite{\isacharunderscore}{\kern0pt}measure\ {\isacharplus}{\kern0pt}\isanewline
\isakeyword{assumes}\ emeasure{\isacharunderscore}{\kern0pt}space{\isacharunderscore}{\kern0pt}{\isadigit{1}}{\isacharcolon}{\kern0pt}\ {\isachardoublequoteopen}emeasure\ M\ {\isacharparenleft}{\kern0pt}space\ M{\isacharparenright}{\kern0pt}\ {\isacharequal}{\kern0pt}\ {\isadigit{1}}{\isachardoublequoteclose}%
}

\usepackage{amsmath}
\usepackage{amssymb}
\usepackage{bbm}
\usepackage{todonotes}

\newcommand{\E}{\mathop{\mathbb{E}}}
\DeclareMathOperator{\parity}{parity}
\DeclareMathOperator{\indicator}{\mathbbm{1}}

\usepackage{algorithm}
\usepackage[noend]{algpseudocode}
\usepackage{cleveref}
\usepackage{booktabs}
\usepackage{adjustbox}

\definecolor{highlightred}{rgb}{.8, 0.0, 0.0}
\definecolor{highlightblue}{rgb}{.14, .21, .868}
\definecolor{highlightgreen}{rgb}{.2,.4,.1}

\newcommand{\redc}[1]{\textcolor{highlightred}{#1}}
\newcommand{\bluec}[1]{\textcolor{highlightblue}{#1}}
\newcommand{\greenc}[1]{\textcolor{highlightgreen}{#1}}

\newcommand{\ApproxMC}{\ensuremath{\mathsf{ApproxMC}}\xspace}

\newcommand{\ApproxMCGen}{\ensuremath{\mathsf{ApproxMCCert}}\xspace}

\newcommand{\ApproxMCCert}{\ensuremath{\mathsf{ApproxMCCert}}\xspace}
\newcommand{\ApproxMCCore}{\ensuremath{\mathsf{ApproxMCCore}}\xspace}
\newcommand{\CertCheck}{\ensuremath{\mathsf{CertCheck}}}

\newcommand{\CryptoMiniSat}{\texttt{CryptoMiniSat}\xspace}
\newcommand{\MiniSat}{\texttt{MiniSat}\xspace}
\newcommand{\FRATrs}{\texttt{FRAT-rs}\xspace}
\newcommand{\FRATxor}{\texttt{FRAT-xor}\xspace}
\newcommand{\cakelpr}{\texttt{cake\_lpr}\xspace}
\newcommand{\cakexlrup}{\texttt{cake\_xlrup}\xspace}
\newcommand{\Tbuddy}{{\texttt{TBUDDY}}\xspace}

\newcommand{\thresh}{\ensuremath{\mathsf{thresh}}}
\newcommand{\BoundedSAT}{\ensuremath{\mathsf{BoundedSAT}}}
\newcommand{\emptyList}{\ensuremath{\mathsf{emptyList}}}
\newcommand{\iter}{\ensuremath{\mathsf{iter}}}
\newcommand{\solCount}{\ensuremath{\mathsf{nSols}}}
\newcommand{\computeIter}{\ensuremath{\mathsf{computeIter}}}
\newcommand{\AddToList}{\ensuremath{\mathsf{AddToList}}}
\newcommand{\FindMedian}{\ensuremath{\mathsf{FindMedian}}}

\newcommand{\FindM}{\ensuremath{\mathsf{FindM}}}

\newcommand{\satisfying}[1]{\ensuremath{\mathsf{sol({#1})}}}
\newcommand{\proj}[2]{\ensuremath{\mathsf{{#1}_{\downarrow#2}}}}

\renewcommand{\satisfying}[1]{\ensuremath{\mathsf{sol}({#1})}}
\renewcommand{\proj}[2]{\ensuremath{{#1}_{\mathsf{\downarrow}#2}}}

\newcommand{\Vars}{\ensuremath{\mathsf{Vars }}}

\newcommand{\PP}{\ensuremath S}

\newcommand{\CaDiCaL}{\ensuremath{\mathsf{CaDiCaL}}\xspace}
\newcommand{\minisatxor}{\ensuremath{\mathsf{MiniSatXOR}}+\ensuremath{\mathsf{pbp}}}
\newcommand{\veripb}{\ensuremath{\mathsf{VeriPB}}\xspace}
\newcommand{\CadCake}{\ensuremath{\mathsf{CaDiCaL}}+\ensuremath{\mathsf{lrat}}}
\newcommand{\cmscake}{\ensuremath{\mathsf{CMS}}+\ensuremath{\mathsf{frat}}-\ensuremath{\mathsf{xor}}}
\newcommand{\cmsbuddy}{\ensuremath{\mathsf{CMS}}+\ensuremath{\mathsf{tbuddy}}}

\newcommand*\circled[1]{\tikz[baseline=(char.base)]{
            \node[shape=circle,draw, minimum size=3.5mm,inner sep=0pt] (char) {#1};}}

\makeatletter
\def\orcidID#1{\href{http://orcid.org/#1}{\protect\raisebox{-1.25pt}{\protect\includegraphics{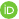}}}}
\makeatother

\begin{document}

\title{Formally Certified Approximate Model Counting\thanks{The first two authors contributed equally.}}
\author{
  Yong Kiam Tan\inst{1}\Envelope\orcidID{0000-0001-7033-2463} \and
  Jiong Yang\inst{2}\orcidID{0000-0002-8356-6637}\and
  Mate Soos\inst{2}\orcidID{0000-0002-7355-881X}\and\\
  Magnus O. Myreen\inst{3}\orcidID{0000-0002-9504-4107} \and
  Kuldeep S. Meel\inst{4}\orcidID{0000-0001-9423-5270}
}

\authorrunning{Y. K. Tan et al.}

\institute{
  Institute for Infocomm Research (I$^2$R), A*STAR, Singapore\\
  \email{tanyk1@i2r.a-star.edu.sg}
\and
  National University of Singapore, Singapore\\
  \email{jiong@comp.nus.edu.sg}\quad
  \email{soos.mate@gmail.com}
\and
  Chalmers University of Technology, Gothenburg, Sweden\\
  \email{myreen@chalmers.se}
\and
  University of Toronto, Toronto, Canada\\
  \email{meel@cs.toronto.edu}
}
\maketitle              %
\begin{abstract}
Approximate model counting is the task of approximating the number of solutions
to an input Boolean formula. The state-of-the-art approximate model counter for
formulas in conjunctive normal form~(CNF), {\ApproxMC}, provides a scalable means
of obtaining model counts with \emph{probably approximately correct}
(PAC)-style guarantees. Nevertheless, the validity of {\ApproxMC}'s approximation
relies on a careful theoretical analysis of its randomized algorithm and the
correctness of its highly optimized implementation, especially the latter's
stateful interactions with an incremental CNF satisfiability solver
capable of natively handling parity (XOR) constraints.

We present the first certification framework for approximate model counting
with formally verified guarantees on the quality of its output approximation.
Our approach combines:
\begin{inparaenum}[(i)]
\item a \emph{static}, once-off, formal proof of the algorithm's PAC guarantee
in the {\isabellehol} proof assistant; and
\item \emph{dynamic}, per-run, verification of {\ApproxMC}'s calls to an
external CNF-XOR solver using proof certificates.
\end{inparaenum}
We detail our general approach to establish a rigorous connection between
these two parts of the verification, including our blueprint for turning the
formalized, randomized algorithm into a verified proof checker, and our design
of proof certificates for both {\ApproxMC} and its internal CNF-XOR solving
steps.
Experimentally, we show that certificate generation adds little overhead to an
approximate counter implementation, and that our certificate checker is able to
fully certify $84.7\%$ of instances with generated certificates when given the
same time and memory limits as the counter.
\end{abstract}


\section{Introduction}\label{sec:introduction}

State-of-the-art automated reasoning solvers are critical software systems used
throughout formal methods. However, even skilled and trusted developers of such
tools can inadvertently introduce errors.  Two approaches have evolved to
provide assurances that automated reasoning tools behave as intended. The first
involves the use of theorem provers to formally verify the correctness of
solver implementations~\cite{DBLP:conf/nfm/Fleury19,DBLP:conf/cpp/KanLRS22}.
This approach guarantees correct outputs for all inputs, but struggles to scale
to complex systems such as SAT solvers.  The second approach is based on
\emph{certifying algorithms}~\cite{DBLP:journals/csr/McConnellMNS11}, where
a solver is required to produce a certificate alongside its
output~\cite{DBLP:journals/jar/BarbosaBFF20,DBLP:conf/sat/BryantNAH23,DBLP:conf/aaai/GochtN21,DBLP:conf/fmcad/KaufmannFB20,DBLP:journals/jar/Lammich20,DBLP:journals/sttt/TanHM23,DBLP:conf/sat/WetzlerHH14}.
A \emph{certificate} checker (also called \emph{proof} checker)---which is
often formally verified---then checks the correctness of this certificate,
ensuring that the system's output adheres to the desired specifications. This
latter method has gained significant traction in the SAT solving community,
wherein a SAT solver either returns a satisfying assignment that is easy to
check through evaluation or a proof of unsatisfiability as a
certificate~\cite{DBLP:conf/sat/WetzlerHH14}.  However, neither of these
approaches have been applied to probabilistic systems that rely on
\emph{randomized} algorithms. In fact, McConnell et
al.~\cite{DBLP:journals/csr/McConnellMNS11} argue that randomized algorithms
resist deterministic certification.

In this paper, we propose a hybrid approach that harnesses the power of both
theorem-proving and certificate-based approaches to certify probabilistic
systems. We present our approach on {\ApproxMC}, a probabilistic automated
reasoning system which computes approximate model counts for Boolean formulas.
Model counting is a fundamental problem in computer science that serves as a
key component in a wide range of applications including
control improvisation~\cite{GVF22},
network reliability~\cite{DMPV17,V79},
neural network verification~\cite{BSSM+19},
probabilistic reasoning~\cite{CFMS+14,EGSS13,R96,SBK05},
and so on. Therefore, it is crucial that the results
computed by an approximate model counter, such as {\ApproxMC}, can be trusted.

Two key questions must be tackled by our approach.  First, what
does it mean to trust a \emph{random} run of {\ApproxMC}? Here, we propose a
{\em verification modulo randomness} approach, i.e., our certification results
are modulo a trusted random bit generator.  Second, how do we handle the huge
volume of (incremental) CNF-XOR satisfiability solver calls which are tightly
integrated in {\ApproxMC}~\cite{DBLP:conf/cav/SoosGM20,DBLP:conf/aaai/SoosM19}?
Here, we design the certificate format to require only the results of solver
calls that are crucial for {\ApproxMC}'s correctness. In particular, {\ApproxMC} makes
$\mathcal{O}(\varepsilon^{-2} \cdot \log n \cdot \log \delta^{-1} )$ many calls to its solver, where $n$ is the number of
(projected) variables of the formula, $\varepsilon$ is the tolerance parameter, and $\delta$ is the confidence parameter (see~\rref{sec:background} for definitions); our crucial insight is that to certify {\ApproxMC}, we only need to
check the correctness of $\mathcal{O}(\log \delta^{-1} )$ UNSAT calls, which is independent of $n$. We then observe that existing CNF-XOR UNSAT checkers
fail to scale to formulas that are handled by {\ApproxMC}. To this end, we adapt existing
solving and verified proof checking pipelines to natively support proof
certificates for CNF-XOR unsatisfiability.  With this design, our framework is
able to independently check certificates generated by a state-of-the-art (but
untrusted) implementation of {\ApproxMC}, with \emph{all} of the latter's
optimizations enabled.  Overall, the key idea is to combine a \emph{static},
once-off, formal proof of the algorithm's correctness guarantee in
{\isabellehol}~\cite{DBLP:books/sp/NipkowPW02,DBLP:journals/jar/Paulson89} with
\emph{dynamic}, per-run, certification of {\ApproxMC}'s calls to an external
CNF-XOR solver.

In summary, our contributions are as follows:
\begin{enumerate}
	\item An abstract specification of {\ApproxMC} and a formal proof of its probably
	approximately correct (PAC) guarantee in {\isabellehol} (\rref{subsec:abstractspec}).
  \item A refinement of the abstract specification to a concrete certificate
  format and checker implementation for {\ApproxMC}
  (Sections~\ref{subsec:certchecker} and~\ref{subsec:approxmccert}).
  \item Updates to various tools to realize a formally verified proof checking
  pipeline with native support for CNF-XOR unsatisfiability
  (\rref{subsec:cnfxorunsat}).
  \item Empirical evaluation of the framework on an extensive suite of
  model counting benchmarks to demonstrate its practical utility
  (\rref{sec:experiments}).
\end{enumerate}

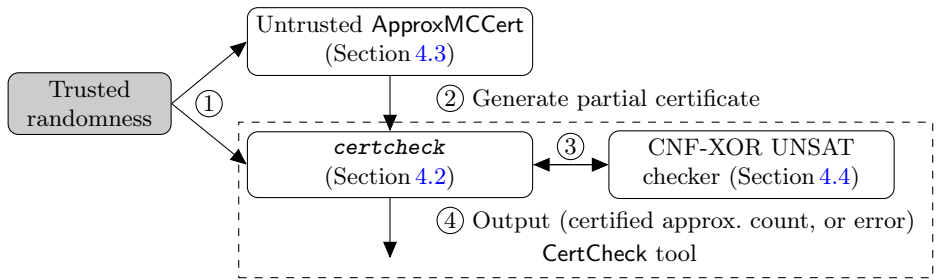
\begin{figure}[t]
	\centering

	\tikzstyle{block} = [rectangle, draw,
	text width=11em, text centered, rounded corners, minimum height=2em, minimum width=5em]
	\tikzstyle{block2} = [rectangle, draw,fill=black!20,
	text width=6.0em, text centered, rounded corners, minimum height=2em, minimum width=5em]

  \tikzset{edge/.style = {->,> = triangle 45}}
	\tikzset{edge2/.style = {<->,> = triangle 45}}
	\begin{tikzpicture}[auto]

		\node [block2] (coins) {Trusted \\ randomness};
		\node [block,above right=-0.05cm and 1cm of coins] (impl) {Untrusted \ApproxMCCert\\(\rref{subsec:approxmccert})};
		\node [block,below right=-0.05cm and 1cm of coins] (cert) {\isa{certcheck}\\(\rref{subsec:certchecker})};
		\node [block,right=1cm of cert] (cnfxor) { CNF-XOR UNSAT\\ checker (\rref{subsec:cnfxorunsat})};

    \node [below = 0.8cm of cert] (concl) {};
    \node [right = 0.2cm of coins] (step1) {\circled{1}};
    \node [above right =-0.5cm and 0.2cm of cert] (step3) {\circled{3}};
    \node [below right = 0.0cm and -1.4cm of impl] {\circled{2} Generate partial certificate};
    \node [below right = 0.0cm and -1.4cm of cert] (outlab) {\circled{4} Output (certified approx. count, or error)};
    \node [below right = 0.5cm and 0cm of cert] (impllab) {{\CertCheck} tool};

		\draw[edge] (coins.east) to (impl.west);
		\draw[edge] (coins.east) to (cert.west);
		\draw[edge] (impl) to (cert);
		\draw[edge] (cert) to (cnfxor) ;
		\draw[edge] (cert) to (concl);
		\draw[edge2] (cert) to (cnfxor);

    \node[draw,dashed,fit=(cert) (impllab) (outlab) (cnfxor)] {};

	\end{tikzpicture}
	\caption{The certified approximate model counting workflow.}
  \label{fig:workflow}
\end{figure}

Our workflow for certified approximate model counting is shown
in~\rref{fig:workflow}.
In step \circled{1}, it uses a trusted external tool to generate uniform random
bits which are handed to an \emph{untrusted} certificate generator
{\ApproxMCCert} and to the verified certificate checker {\CertCheck} (extracted
from {\isabellehol});
the random bits are used identically by {\ApproxMCCert} and {\CertCheck} to
generate random XOR constraints as part of the counting algorithm.
For step \circled{2}, {\ApproxMCCert} generates a \emph{partial} certificate
which is subsequently checked in step \circled{3} by
{\CertCheck}; the certificate is partial because it does not contain
CNF-XOR unsatisfiability proofs.
Instead, {\CertCheck} calls an external CNF-XOR unsatisfiability checking
pipeline (with verified proof checking in
CakeML~\cite{DBLP:conf/popl/KumarMNO14,DBLP:journals/sttt/TanHM23}).
In the final step \circled{4}, an approximate model count is returned upon
successful certification.

As part of our commitment to reproducibility, all code and proofs have been made available
with a permissive open-source
license~\cite{approxmccert,fratxor,Approximate_Model_Counting-AFP}.
\subsubsection{Impact.} Although our main objective was to enhance end-user
trust in answers to their counting queries, undertaking this project led to
unexpected benefits that are worth highlighting.  While modifying {\ApproxMC}'s
underlying solver, {\CryptoMiniSat}~\cite{DBLP:conf/sat/SoosNC09}, to emit
certificates (\rref{subsec:cnfxorunsat}), a bug in {\CryptoMiniSat}'s XOR manipulation system was discovered.
The bug was introduced during the development of part of the BIRD
system~\cite{DBLP:conf/aaai/SoosM19} that keeps \emph{all} XOR constraints'
clausal versions (as well as their compact XOR versions) in-memory at all
times. This allows a substantial level of interaction between XOR and clausal
constraints.  However, it also led to large overhead in terms of the often
hundreds of thousands of clauses needed to encode the XORs in their clausal
form.  The compromise made by the developers was to detach the clausal
representation of XORs from the watchlists.  However, that seemed to have led
to a level of complexity that both allowed the bug to occur, and more
importantly, made it impossible to discover via {\CryptoMiniSat}'s standard
fuzzing pipeline.  Our version of {\CryptoMiniSat} fixes this by not keeping
around a clausal encoding of all XORs, instead introducing (and deleting) them
whenever needed for the proof.

Furthermore, we have also found minor flaws in the
theoretical analysis of {\ApproxMC} (see discussion of \isa{events\_prob}) and
in the implementation, e.g., the sampling of random bits was slightly biased,
and an infinite loop could be triggered on certain random seeds. None of these
bugs were known to the authors of {\ApproxMC} or were previously reported by
users of the tool. All of these
issues have been fixed and upstreamed to their tools' respective
codebases.

\section{Related Work}\label{sec:related}

This discussion is focused on formally verified algorithms and
proof checkers. Readers are referred to Chakraborty et
al.~\cite{DBLP:series/faia/ChakrabortyMV21} and references therein for related
literature on approximate model counting.

\paragraph{Certified Model Counting.}
Prior research on certificate-based approaches focuses on deterministic
methods in model counting.  Prior work on certified \emph{exact} model
counting focuses either on the development of proofs, such as
MICE~\cite{FHR22} and CPOG~\cite{DBLP:conf/sat/BryantNAH23}, along with
their respective toolchains, or on analyzing the complexity of the
proof system~\cite{BHS23}.  Some efforts have been directed toward
certifying deterministic approximate counting algorithms which,
however, require access to a $\mathrm{\Sigma^{P}_2}$ oracle and did not
yield practical implementations~\cite{MCA24}.  Our work develops the
first certification framework for randomized approximate model
counting.

\paragraph{Formalization of Randomized Algorithms.} Various randomized
algorithms have been formally analyzed in {\isabellehol}, including randomized
quicksort, random binary tree data
structures~\cite{DBLP:journals/jar/EberlHN20}, and approximation of frequency
moments in data
streams~\cite{DBLP:conf/itp/Karayel22,Frequency_Moments-AFP,Median_Method-AFP,Universal_Hash_Families-AFP}.
These prior efforts as well as ours, all build upon the foundations for measure
and probability theory in
{\isabellehol}~\cite{DBLP:conf/esop/EberlHN15,DBLP:conf/itp/HolzlLT15}.
Properties of approximate membership query structures (including Bloom filters)
have been verified in Coq~\cite{DBLP:conf/cav/GopinathanS20}.  Pioneering work
on formal verification of randomized algorithms, including the Miller-Rabin
primality test, was carried out by Joe Hurd in HOL4~\cite{UCAM-CL-TR-566}. A
common objective of these prior efforts, and that of ours, is to put the
guarantees of randomized algorithms on formal foundations.

\paragraph{Verified Proof Checking.}
Formally verified proof checkers have been developed for several
(deterministic) algorithms and theories, such as the CNF unsatisfiability
checkers used by the SAT
community~\cite{DBLP:conf/itp/HeuleHKW17,DBLP:journals/jar/Lammich20,DBLP:journals/sttt/TanHM23}.
Within {\isabellehol}, the Past\`eque tool~\cite{DBLP:conf/fmcad/KaufmannFB20}
checks proofs in the practical algebraic calculus, which can be used to
validate algebraic reasoning; the CeTA tool~\cite{DBLP:conf/tphol/ThiemannS09}
is based on an extensive library of results for certifying properties of
rewriting systems; and the LEDA project developed specialized proof checkers
for graph algorithms~\cite{DBLP:conf/mfcs/AbdulazizMN19}.
 CoqQFBV~\cite{DBLP:conf/cav/ShiFLTWY20} is similar in design to our
 approach in that a higher-level Coq-generated tool for verified
 bit-blasting is used in concert with a lower-level verified proof
 checker for CNF formulas.

\paragraph{CNF-XOR Unsatisfiability Checking.} Given {\ApproxMC}'s reliance on
CNF-XOR formulas, certification of CNF-XOR unsatisfiability emerged as a key
challenge in our work. To this end, we provide a brief overview of three prior
state-of-the-art approaches for certified CNF-XOR reasoning.
\begin{enumerate}
  \item\label{item:bdd} The first approach uses proof generation and certification of XOR
  reasoning based on Binary Decision Diagrams
  (BDDs)~\cite{DBLP:journals/corr/abs-2304-04292}. It uses
  {\CryptoMiniSat}~\cite{DBLP:conf/sat/SoosNC09}, a SAT solver specifically
  made to work on CNF-XOR instances and
  {\Tbuddy}~\cite{DBLP:conf/fmcad/Bryant22} to produce FRAT proof certificates~\cite{DBLP:journals/lmcs/BaekCH22} for
  {\CryptoMiniSat}'s XOR reasoning;
  {\FRATrs}~\cite{DBLP:journals/lmcs/BaekCH22} is used as the elaboration
  backend and a verified LRAT proof
  checker~\cite{DBLP:conf/itp/HeuleHKW17,DBLP:journals/sttt/TanHM23} can be
  used to check the elaborated proofs.
  \item\label{item:pbp} The second approach, due to Gocht and
  Nordstr\"om~\cite{DBLP:conf/aaai/GochtN21},  relies on pseudo-Boolean reasoning
  and its associated proof system to justify both CNF and parity
  reasoning. This approach was demonstrated on MiniSat equipped with an XOR
  reasoning engine, with \veripb as a proof checker; pseudo-Boolean proofs are
  also supported by a verified proof checker~\cite{subgraph-aaai24}.
  \item\label{item:cdcl} The third approach is to rely on the standard SAT solvers accompanied
  with standard CNF proof formats and (verified)
  checkers~\cite{DBLP:conf/itp/HeuleHKW17,DBLP:journals/jar/Lammich20,DBLP:journals/sttt/TanHM23}.
\end{enumerate}

\section{Background}\label{sec:background}

This section gives a brief introduction to {\ApproxMC}
(\rref{sec:background-appmc}) and to theorem-proving in {\isabellehol}
(\rref{sec:background-isabelle}).

\subsection{Approximate Model Counting}\label{sec:background-appmc}

Given a Boolean formula $F$, the \emph{model counting} problem is to calculate
the number of models (also called \emph{solutions} or \emph{satisfying assignments}) of $F$. Model counting is known to be \#P-complete, and
therefore has been a target of sustained interest for randomized approximation
techniques over the past four decades.  The current state-of-the-art approximate  approach,
{\ApproxMC}~\cite{DBLP:conf/ijcai/ChakrabortyMV16}, is a hashing-based
framework that relies on reducing the model counting problem to SAT
queries, which are handled by an underlying solver.  Importantly, {\ApproxMC}
is a \emph{probably approximately correct} (PAC) projected model counter, i.e.,
it takes in a
formula $F$, a projection set $\PP \subseteq \Vars(F)$, a tolerance parameter $\varepsilon >
0$, and a confidence parameter $\delta \in (0,1]$, and
returns a count $c$ satisfying the PAC guarantee: $\Pr
\left[\frac{|\proj{\satisfying{F}}{\PP}|}{1+\varepsilon} \le c \le (1+\varepsilon)|\proj{\satisfying{F}}{\PP}|\right] \geq 1 - \delta$, where $|\proj{\satisfying{F}}{\PP}|$ denotes the number of the solutions of $F$ projected on $\PP$.

\begin{algorithm}[t]
	\caption{\ApproxMC$(F, \PP, \varepsilon, \delta)$}\label{alg:appmc}

	\begin{algorithmic}[1]
		\State $\thresh \leftarrow 9.84 \left(1+\frac{\varepsilon}{1+\varepsilon}\right)\left(1+\frac{1}{\varepsilon}\right)^2$ \label{ln:thresh}
		\State $Y \leftarrow \BoundedSAT(F, \PP, \thresh)$\label{ln:appmc-bounded-sat}
		\If{$(|Y| < \thresh)$} \Return $|Y|$\EndIf
		\State $t \leftarrow \computeIter(\delta)$\Comment probability amplification using the median method \label{ln:iter}\\
		$C\leftarrow\emptyList, \iter \leftarrow 0$\label{ln:init}
		\Repeat	\label{ln:repeat-begin}
		\State $\iter \leftarrow \iter + 1$
		\State $\solCount \leftarrow \ApproxMCCore(F, \PP, \thresh)$\label{ln:invoke-core}
		\State $\AddToList(C,\solCount)$
		\Until{$(\iter \ge t)$}\label{ln:repeat-end}
		\State \Return $\FindMedian(C)$\label{ln:find-median}
	\end{algorithmic}
\end{algorithm}
\begin{algorithm}[t]
	\caption{\ApproxMCCore$(F,\PP,\thresh)$}\label{alg:appmc-core}

	\begin{algorithmic}[1]
		\State Choose $|\PP|-1$ random XOR constraints $X = (X_1,\dots,X_{|\PP|-1})$ over $S$
		\State $m \leftarrow \FindM(F, \PP, X, \thresh)$\label{ln: estimate m} \Comment search for $m \in \{1,\dots,|\PP|\}$ using \BoundedSAT \label{ln:log-search}
		\If{$(m \ge |\PP|)$} \Return $(2^m\times 1)$\Comment dummy value for failed round\EndIf \label{ln:appmc-core-bad}
		\State $c \leftarrow \BoundedSAT\left(F\land X_1 \land \dots \land X_m , \PP, \thresh\right);$\label{ln:appmc-core-bounded-sat}
		\State \Return $(2^m\times c)$\label{ln:core-return}
	\end{algorithmic}
\end{algorithm}

An outline of {\ApproxMC} is shown in Algorithms~\ref{alg:appmc}
and~\ref{alg:appmc-core}. At a high level, the key idea of {\ApproxMC} is to
partition the set of solutions into small cells of roughly equal size by
relying on the power of XOR-based hash
families~\cite{DBLP:conf/ijcai/ChakrabortyMV16,DBLP:conf/nips/GomesSS06}, then
randomly picking one of the cells and enumerating all the solutions in the
chosen small cell up to a threshold $\thresh$ via calls to $\BoundedSAT(F, \PP,
\thresh)$. The estimated count is obtained by scaling the number of solutions
in the randomly chosen cell by the number of cells, and the success probability
of this estimation is amplified to the desired level by taking the median
result from several trials.

Syntactically, the solution space partition and random cell selection is
accomplished by introducing randomly generated XOR constraints of the form
$(\bigoplus_{y \in Y} y) = b$ for a random subset $Y \subseteq \PP$ and random
bit $b$. A crucial fact about random XOR constraints exploited by {\ApproxMC}
is their 2-universality when viewed as a hash family on assignments---briefly,
given any two distinct Boolean assignments over the variable set $\PP$, the
probability of each one satisfying a randomly chosen XOR constraint is
independent and equal to $\frac{1}{2}$.

Accordingly, the $\BoundedSAT$ queries made in Algorithms~\ref{alg:appmc}
and~\ref{alg:appmc-core} are conjunctions of the input formula and random XOR
constraints, i.e., CNF-XOR formulas. The current implementation of {\ApproxMC}
relies on {\CryptoMiniSat} for its ability to handle CNF-XOR formulas
efficiently and
incrementally~\cite{DBLP:conf/cav/SoosGM20,DBLP:conf/aaai/SoosM19}.
Furthermore, the real-world implementation also relies on three key
optimizations.  \begin{inparaenum}[\bf (1)] \item The search for the correct
value of $m$ in Algorithm~\ref{alg:appmc-core} ($\FindM$) combines a linear
neighborhood search, a galloping search, and a binary
search~\cite{DBLP:conf/ijcai/ChakrabortyMV16}.  \item The underlying SAT solver
is used as a library, allowing to solve under a set of assumptions, a technique
introduced as part of MiniSat~\cite{10.1007/978-3-540-24605-3_37}. This allows
the solver to keep learned lemmas between subsequent calls to {\tt solve()},
significantly improving solving speed, which is especially helpful for proving
unsatisfiability.  \item To improve the speed of finding satisfying
assignments, a solution cache of past solutions is
retained~\cite{DBLP:conf/cav/SoosGM20} which is especially helpful when the
optimal number of \bluec{XORs} to add is N, but N+1 have been added and were
found to be too much. In these cases, all solutions that are valid for N+1
\bluec{XORs} are also solutions to N \bluec{XORs} and can be reused.
\end{inparaenum}

\subsection{Formalization in Isabelle/HOL}\label{sec:background-isabelle}

\subsubsection{Notation.}
All \isabellehol syntax is typeset in \isa{typewriter} font with boldface Isar
\isacommand{keywords}; \isa{\isasymAnd} and \isa{\isasymLongrightarrow} are the
universal quantifier and implication of Isabelle's metalogic, respectively.
Type variables are written as \isa{'a, 'b}. The type of (total) functions from
\isa{'a} to \isa{'b} is written as \isa{'a~\isasymRightarrow~'b}, and the type
of partial functions, which are only defined on some elements of type \isa{'a},
is \isa{'a~\isasymrightharpoonup~'b}. For clarity, we often annotate terms with
their type using the notation \isa{term :: type}. For types such as reals,
integers, or natural numbers, the interval from \isa{i} to \isa{j} (inclusive)
is written as \isa{\{i..j\}}; the same interval except endpoint \isa{j} is
\isa{\{i..<j\}}.
More comprehensive introductions can be found in standard
references~\cite{DBLP:journals/jar/Ballarin14,DBLP:books/sp/NipkowPW02}.

\subsubsection{Locales and Probability.}
\isabellehol is equipped with \emph{locales}~\cite{DBLP:journals/jar/Ballarin14}, a system of user-declared modules
consisting of syntactic parameters, assumptions on those parameters, and
module-specific theorems. These modules can be instantiated and inherited,
giving users a powerful means of managing mathematical relationships.
The following snippet, taken from the \isabellehol standard library, shows an
example \isacommand{locale} declaration for probability spaces followed by
an \isacommand{interpretation} command claiming that the measure space
associated with any probability mass function (PMF) \isa{p} is a probability space~\cite{DBLP:conf/itp/HolzlLT15}.

\begin{isabelle}\begin{mdframed}[backgroundcolor=black!10,linecolor=black!10]
\localeprobspace\\
\dots\\\localemeasurepmf
\end{mdframed}\end{isabelle}

Thanks to the locale interpretation, all definitions and theorems associated
with probability spaces can be used with PMFs. For example, the probability of
an event \isa{A :: 'a set} occurring under \isa{p} is \isa{measure\_pmf.prob p
A}.
The support of PMF \isa{p} is \isa{set\_pmf p}, which is \isa{finite}
for all PMFs considered in this work.

\section{Approximate Model Counting in Isabelle/HOL}\label{sec:formalization}

This section outlines our formalization of {\ApproxMC} in {\isabellehol} and
its verified certificate checker implementation.  The proof follows a
refinement-based approach, starting with an abstract mathematical specification
of {\ApproxMC}, where its probabilistic approximation guarantees can be
formalized without low-level implementation details getting in the way
(\rref{subsec:abstractspec}). Then, the abstract specification is progressively
concretized to a verified certificate checker which we call {\CertCheck}
(\rref{subsec:certchecker}) and we extend {\ApproxMC} to {\ApproxMCCert}, a
certificate-generating counter (\rref{subsec:approxmccert}). As part of
\CertCheck, we also built a native CNF-XOR unsatisfiability checker, which is
external to {\isabellehol}, but is also based on formally verified proof
checking (\rref{subsec:cnfxorunsat}).

\subsection{Abstract Specification and Probabilistic Analysis}\label{subsec:abstractspec}

Throughout this section, the type \isa{'a} abstracts the syntactic
representation of variables. For example, in the DIMACS CNF
format,
variables
are represented with positive numbers, while in other settings, it may be more
convenient to use strings as variable names. A solution (or model)
\isa{w :: 'a~\isasymRightarrow~bool} is a Boolean-valued function on variables and a
projection set \isa{S :: 'a set} is a (finite) set of variables. The main
result of this section is formalized in a locale with two parameters
\isa{sols}, \isa{enc\_xor}, and an assumption relating the two:
\begin{isabelle}\begin{mdframed}[backgroundcolor=black!10,linecolor=black!10]
\localeappmc
\end{mdframed} \end{isabelle}

Here, type \isa{'fml} abstracts the syntactic representation of formulas,
\isa{sols F} is the set of all solutions of a formula \isa{F}, and
\isa{enc\_xor xor F} is a formula whose set of solutions satisfies both \isa{F}
and the XOR constraint \isa{xor}. An instantiation of the \isa{ApproxMC} locale
would need to provide implementations of \isa{sols}, \isa{enc\_xor} and prove
that they satisfy the latter assumed property.

The PAC theorem for \ApproxMC is formalized as follows:
\begin{isabelle}\begin{mdframed}[backgroundcolor=black!10,linecolor=black!10]
\appmcprob
\end{mdframed} \end{isabelle}

Here, \isa{sz} is the true count of projected solutions, i.e., the cardinality of the set \isa{proj S (sols F)}, interpreted as a real number.
The conclusion says that \isa{approxmc} returns an
\isa{\isasymepsilon}-approximate count \isa{c} with probability at least
\isa{1\ -\ \isasymdelta}. The argument \isa{n} is a user-specifiable minimum
number of iterations of \ApproxMCCore calls inside \ApproxMC; in practice, a sufficient
number of rounds is automatically determined using the median method. Since the
\isa{ApproxMC} locale can be instantiated for \emph{any} Boolean theory in
which XOR constraints can be syntactically encoded, this theorem shows that the
approximate model counting algorithm of Chakrabory et
al.~\cite{DBLP:conf/ijcai/ChakrabortyMV16} works for any such theory.

The rest of this section gives an overview of our proof of
\isa{approxmc\_prob}. Technical differences compared
to the original proofs are discussed in remarks.

\subsubsection{Formalized Analysis of \ApproxMCCore.} For simplicity, we write
\isa{S~\isasymRightarrow~bool} for the type of solutions projected onto set
\isa{S} and \isa{[n]~\isasymRightarrow~bool} for $n$-dimensional bit-vectors,
i.e., the type of Boolean-valued functions on domain $0,1,\dots,n-1$. A hash
function \isa{h ::
(S~\isasymRightarrow~bool)~\isasymRightarrow~([n]~\isasymRightarrow~bool)} maps
projected solutions into $n$-dimensional bit-vectors. Let \isa{W ::
('a~\isasymRightarrow~bool) set} be any set of solutions, such as \isa{sols F}.
Abstractly, \ApproxMCCore is a way of approximating the cardinality of the
projected set \isa{proj S W}, given an \emph{oracle} that can count up to a
specified threshold \isa{thresh} number of solutions. Without loss of
generality, assume \isa{thresh \isasymle\ proj S W} (otherwise, the oracle
returns the exact count).

\begin{remark} The simple type theory of \isabellehol does not support
dependent function types like \isa{S~\isasymRightarrow~bool} and
\isa{[n]~\isasymRightarrow~bool}.  Our formalization represents functions with
type \isa{S~\isasymRightarrow~bool} as partial functions
\isa{'a~\isasymrightharpoonup~bool} along with an assumption that their
function domain is equal to \isa{S}.
\end{remark}

For any fixed bit-vector \isa{\isasymalpha\ :: [card S -
1]~\isasymRightarrow~bool}, the sets of hash functions \isa{T}, \isa{L}, and
\isa{U} used in the analysis are defined as follows, where \isa{card\_slice h
i} counts the number of entries of \isa{w \isasymin\ proj S W} such that the
hash value \isa{h w} agrees with \isa{\isasymalpha} on their first \isa{i}
entries (also called the \emph{i}-th slices).
\begin{isabelle}
\begin{mdframed}[backgroundcolor=black!10,linecolor=black!10]
\TLUdef
\end{mdframed}
\end{isabelle}

For any input hash function \isa{h}, the following \isa{approxcore} function (cf. Algorithm~\rref{alg:appmc-core} Lines 2--5)
finds the first index \isa{m}, if one exists in \isa{[1..<card S]}, where
\isa{h \isasymin\ T m}. It returns the approximate model count as a multiplier
(\isa{2 \isacharcircum\ m}) and cell size (\isa{card\_slice h m}).
The \emph{failure event} \isa{approxcore\_fail} is the set of hash functions \isa{h}
such that \isa{approxcore} returns a non-\isa{(1+\isasymepsilon)}-factor-approximate count.
\begin{isabelle}\begin{mdframed}[backgroundcolor=black!10,linecolor=black!10]
	\appcore
\end{mdframed} \end{isabelle}

The key lemma for \isa{approxcore} (shown with proof sketch below) is that, for
hash functions \isa{h}, which are randomly sampled from an appropriate hash
family \isa{H}, the probability of the aforementioned failure event is bounded
above by $0.36$~\cite{DBLP:conf/ijcai/ChakrabortyMV16}.
The lemma uses \isabellehol's formalization of hash families which
is \emph{seeded}~\cite{DBLP:conf/itp/Karayel22}, i.e., \isa{p} is
a PMF on seeds and \isa{H} is a \isa{2}-universal hash family for
seeds drawn from \isa{p}; \isa{map\_pmf
\isa{{\isacharparenleft}{\kern0pt}{\isasymlambda}s\
w{\isachardot}{\kern0pt}\ H\ w\ s{\isacharparenright}{\kern0pt}}
p} is a PMF which samples a random seed \isa{s} and then returns
the hash function associated with that seed according to the
family \isa{H}.

\begin{isabelle}\begin{mdframed}[backgroundcolor=black!10,linecolor=black!10]
\appcorefailprob
\end{mdframed} \end{isabelle}

\begin{proof}
The proof of \isa{approxcore\_fail\_prob} proceeds via several
sub-lemmas~\cite{DBLP:conf/ijcai/ChakrabortyMV16}, which we discuss inline below.
We first show that an index \isa{mstar} exists with the following properties
(\isakeyword{obtains} is the Isar keyword for existential claims):
\begin{isabelle}\begin{mdframed}[backgroundcolor=black!10,linecolor=black!10]
\mstarexists
\end{mdframed} \end{isabelle}

This is proved by noting that there exists \isa{m} satisfying the
first two properties separately in the finite interval \isa{1,2,\dots,card S -
1}, so there must be an \isa{mstar} satisfying all three properties in that
interval.

Next, the failure event (which is a set of hash functions) is proved to be
contained in the union of four separate events involving \isa{mstar} using
the properties from \isa{mstar\_exists} and unfolding the respective
definitions of \isa{T}, \isa{L}, and \isa{U}:
\begin{isabelle}\begin{mdframed}[backgroundcolor=black!10,linecolor=black!10]
\failuresubset
\end{mdframed} \end{isabelle}

Finally, we bound the probability for each of the four events separately.
\begin{isabelle}\begin{mdframed}[backgroundcolor=black!10,linecolor=black!10]
\eventsprob
\end{mdframed} \end{isabelle}

Lemma \isa{approxcore\_fail\_prob} follows from \isa{failure\_subset},
\isa{events\_prob}, and the union bound on probabilities. \qed
\end{proof}

\begin{remark}
Our implicit construction of \isa{mstar} in \isa{mstar\_exists} avoids an
explicit calculation from \isa{F}, \isa{S} and
\isa{\isasymepsilon}~\cite{DBLP:conf/ijcai/ChakrabortyMV16}, which is more
intricate to analyze.
Additionally, in \isa{events\_prob}, the first bound for \isa{T (mstar-3)}
works only when \isa{{\isasymepsilon}\ {\isasymle}\ {\isadigit{1}}}, an omitted
condition from the pen-and-paper proof~\cite[Lemma
2]{DBLP:conf/ijcai/ChakrabortyMV16}; we also verified a looser bound of \isa{1
/ 10.84} without this condition, but this leads to a weaker overall guarantee
for \ApproxMCCore (which we do not use subsequently).
\end{remark}

\subsubsection{Formalized Analysis of \ApproxMC.}

Random XORs and XOR-based hash families are defined as follows:
\begin{isabelle}\begin{mdframed}[backgroundcolor=black!10,linecolor=black!10]
\randomxor

\xorhashfamily
\end{mdframed} \end{isabelle}

Here, \isa{random\_xor V} is the PMF which samples a pair of a uniformly
randomly chosen subset of the (projection) variables \isa{V} and the outcome of
a fair coin flip; \isa{random\_xors V n} is the PMF that samples \isa{n}
independent XORs according to \isa{random\_xor V}.  Given \isa{card S - 1}
randomly chosen seed \isa{xors}, the associated \isa{xor\_hash} hash function
takes a projected solution \isa{w} to the bit-vector whose bit \isa{i}
indicates whether the \isa{i}-th XOR is satisfied by \isa{w}.

The following definition of \isa{approxmccore} (cf. Algorithm~\rref{alg:appmc-core}) randomly samples \isa{card S\ -\ 1}
XOR constraints over the variables \isa{S} and runs
\isa{approxcore\_xors} (\isa{approxcore} instantiated with XOR-based hash
families using \isa{xor\_hash}).
The top-level function \isa{approxmc} (cf. Algorithm~\rref{alg:appmc}) selects appropriate values for
\isa{thresh} and the number of rounds \isa{t} for amplification using the
median method.
\begin{isabelle}\begin{mdframed}[backgroundcolor=black!10,linecolor=black!10]
\approxmcdef
\end{mdframed} \end{isabelle}

The main result \isa{approxmc\_prob} follows from 2-universality of XOR-based
hash families and the facts that \isa{compute\_thresh} returns a correct value
of \isa{thresh} and \isa{compute\_t} chooses a sufficient number of rounds for
the median method.

\subsubsection{Library Contributions.} We added reusable results
to \isabellehol's probability libraries, such as the Paley-Zigmund inequality
(a concentration inequality used in the analysis of \ApproxMCCore) and a
slightly modified (tighter) analysis of the median method based on the prior
formalization by Karayel~\cite{DBLP:conf/itp/Karayel22,Median_Method-AFP};
the latter modification does not change the asymptotic analysis of the
method but it is needed as \ApproxMC implementations use the tighter
calculation to reduce the number of rounds for success probability amplification.

We also formalized the 3-universality of XOR-based hash
families~\cite{DBLP:conf/nips/GomesSS06}, which implies its 2-universality, as
needed by \ApproxMC.
\iflongversion
The proof used in our formalization is sketched in~\rref{app:threeunivproof}.
\else
The proof is sketched in the online extended version of this paper.
\fi
Our (new) proof is of independent interest as it is purely combinatorial, using
a highly symmetric case analysis which helps to reduce formalization effort
because many cases can be proved using without-loss-of-generality-style
reasoning in \isabellehol.

\subsection{Concretization to a Certificate Checker}\label{subsec:certchecker}

The specification from~\rref{subsec:abstractspec} leaves several
details abstract.  For example, \isa{card\_slice} refers to set
cardinalities and \isa{approxmc} uses a bounded solution
counter as an oracle, neither of which are \emph{a priori} computable terms. This
section gives a concrete implementation strategy where the abstract
details are obtained from certificates generated by an untrusted
external implementation, and checked using verified code. The main
result is formalized in a locale \isa{CertCheck}
with two key extensions compared to \isa{ApproxMC} from~\rref{subsec:abstractspec}:
\begin{inparaenum}[(i)]
\item \label{itm:listswap}the \isa{ApproxMCL} locale, switching from
set-based to computable list-based representations for the projection set and
XORs;
\item \label{itm:checkbansol}the additional locale parameters
\isa{check\_sol} determining whether a formula is satisfied by a
specified assignment, and \isa{ban\_sol} that syntactically blocks a
solution from further consideration.
\end{inparaenum}

\begin{isabelle}\begin{mdframed}[backgroundcolor=black!10,linecolor=black!10]
\localeappmclcert
\end{mdframed} \end{isabelle}

The correctness of the \isa{certcheck} checker (shown below) has two
conjuncts in its conclusion.
In both conjuncts, \isa{f} models an external
(untrusted) implementation returning a certificate and \isa{r} is a random seed
passed to both \isa{f} and \isa{certcheck}. The checker either returns an
error string (\isa{isl}) or a certified count.
The \emph{soundness} guarantee
(left conjunct) says that the probability of the checker returning an incorrect
count (without error) is bounded above by \isa{\isasymdelta}.  Note that for a
buggy counter \isa{f} that always returns an invalid certificate,
\isa{certcheck} returns an error for all random seeds, i.e., it returns
a count (whether correct or not) with probability 0.  Thus, the
\emph{promise-completeness} guarantee (right conjunct) says that \emph{if} the
function \isa{f} is promised to return valid certificates for all seeds
\isa{r}, then the checker returns a correct count with probability \isa{1 -
\isasymdelta}.
\begin{isabelle}\begin{mdframed}[backgroundcolor=black!10,linecolor=black!10]
\appmclcertprob
\end{mdframed} \end{isabelle}

Additional differences in \isa{certcheck\_prob} compared to \isa{approxmc\_prob} are:
\begin{inparaenum}[(i)]
\setcounter{enumi}{2}
\item \label{itm:checkunsat}the oracle function \isa{check\_unsat}, which is
    assumed to be an interface to an external unsatisfiability checker;
\item \label{itm:certificates}the additional certificate arguments \isa{m0} and
    \isa{ms}; and
\item \label{itm:eager}the eager sampling of XORs using random
    bits (\isa{random\_seed\_xors}), compared to \isa{approxmc} which samples
    lazily.
\end{inparaenum}

\begin{remark}
Note that \isa{ban\_sol} and \isa{check\_sol} are locale parameters with
assumptions that must be \emph{proven} when \isa{CertCheck} is
instantiated to a Boolean theory; in contrast, \isa{check\_unsat} appears
as an \emph{assumption}. The pragmatic reason for this difference is that
\isa{ban\_sol} and \isa{check\_sol} can be readily implemented in
\isabellehol with decent performance. In contrast, developing efficient
verified \emph{unsatisfiability} proof checkers and formats, e.g., for
CNFs, is still an active area of
research~\cite{DBLP:journals/lmcs/BaekCH22,DBLP:conf/itp/HeuleHKW17,DBLP:journals/jar/Lammich20,DBLP:journals/sttt/TanHM23}.
Leaving \isa{check\_unsat} outside the scope of \isabellehol allows us to rely
on these orthogonal verification efforts (as we do in~\rref{subsec:cnfxorunsat}).
\end{remark}

\begin{figure*}[t]
\renewcommand{\ttdefault}{pcr}
\centering
\begin{minipage}[t]{.23\textwidth}
\centering
{Input formula}
\begin{Verbatim}[frame=single,fontsize=\small]
p cnf 10 7
1 2 3 4 5 0
6 7 8 9 10 0
-1 -6 0
-2 -7 0
-3 -8 0
-4 -9 0
-5 -10 0
\end{Verbatim}
{Approximate count}
\begin{Verbatim}[frame=single,fontsize=\small]
...
s mc 184
\end{Verbatim}
\end{minipage}
\quad
\begin{minipage}[t]{.72\textwidth}
\centering
{\isa{certcheck} partial certificate file}
\begin{Verbatim}[frame=single,commandchars=\\\{\},fontsize=\small]
0   // initial m0
\redc{73}  // \redc{number} and \bluec{list of solutions}
\bluec{-1 2 -3 -4 -5 6 -7 -8 -9 -10 0}
\bluec{...}
\bluec{1 -2 -3 4 5 -6 7 -8 -9 -10 0}
2   // round 1 value of m
\redc{73}  // \redc{number} and \bluec{list of solutions}
\bluec{...} //   \bluec{after adding m - 1 XORs}
\redc{51}  // \redc{number} and \bluec{list of solutions}
\bluec{...} //   \bluec{after adding m XORs}
    //   \greenc{*UNSAT after excluding 51 solutions}
    //   \greenc{checked by external pipeline}
2   // round 2 value of m
    // ... repeat for t rounds ...
\end{Verbatim}
\end{minipage}
\caption{An example pigeon-hole formula (2 pigeons, 5 holes, 180
    solutions) in DIMACS format and a valid certificate for the checker at
    $\varepsilon=0.8$ and $\delta=0.2$ ($\isa{thresh}=73, \isa{t}=9$). The
    certificate is shown with colored comments and with redundant spaces added
    for clarity. In clauses, the negative (resp. positive) integers
    are negated (resp. positive) literals, with a 0 terminator;
    solutions are lists of literals assigned to true.
    Part of the certificate (marked with \greenc{\texttt{*}}) is checked with
    an external UNSAT proof checking pipeline.
    }
\label{fig:appmccertformat}
\end{figure*}

\subsubsection{From \isa{approxmc} to \isa{certcheck}.}
We briefly list the steps in transporting the PAC guarantee from
\isa{approxmc} to \isa{certcheck}, with reference to the differences labeled
(\rref{itm:listswap})--(\rref{itm:eager}) above.
The proof follows a sequence of small refinement steps which are individually
straightforward as they do not involve significant probabilistic reasoning.
First, cf.~(\rref{itm:eager}), a variant of \isa{approxmc} is formalized where
all XORs are eagerly sampled upfront, as opposed to lazily at each call to
\isa{approxmccore}. Without loss of generality, it suffices to sample \isa{t}
$\times$ \isa{(card S\ -\ 1)} XORs.
Next, cf.~(\rref{itm:listswap}), the representations are swapped to executable
ones, e.g., the projection set is represented as a list \isa{S} of distinct
elements. Accordingly, the left-hand side of each XOR is represented as a list
of \isa{length S} bits, where the \isa{i}-th bit indicates whether the
\isa{i}-th entry of \isa{S} is included in the XOR. Note that it suffices to
sample \isa{t} $\times$ \isa{(card S\ -\ 1)} $\times$ \isa{(card S\ +\ 1)} bits
for \ApproxMC.
Finally, cf.~(\rref{itm:certificates}), \emph{partial certificates} are
introduced. The key observation is that the final value of \isa{m} in
\isa{approxcore} from~\rref{subsec:abstractspec} can be readily
\emph{certified} because it is the first entry where adding \isa{m} XORs causes
the solution count to fall below \isa{thresh}---the solution count is
monotonically decreasing as more XORs are added. Thus, for a claimed value of
\isa{m} it suffices to check, cf.~(\rref{itm:checkbansol}) and
(\rref{itm:checkunsat}) that the following three conditions hold.
\begin{inparaenum}[\bf (1)]
\item Firstly, \isa{1 \isasymle\ m \isasymle\ card S - 1}.
\item Secondly, the solution count after adding \isa{m - 1} XORs reaches or exceeds
    \isa{thresh}, which can be certified (\isa{check\_sol}) by a list of
    solutions of length at least \isa{thresh}, which are distinct after
    projection on \isa{S}.
\item Thirdly, if \isa{m < card S - 1}, then the solution count after adding \isa{m}
    XORs is below \isa{thresh}, which can be certified (\isa{check\_sol}) by a
    list of solutions of length below \isa{thresh}, which are distinct
    after projection, and where the formula after excluding all those projected solutions (\isa{ban\_sol})
    is unsatisfiable (\isa{check\_unsat}).
\end{inparaenum}
An example partial certificate is shown in~\rref{fig:appmccertformat}.
Note that we call these \emph{partial} certificates because of the
reliance on an external pipeline for checking unsatisfiability, as illustrated in the example.

\subsubsection{Code Extraction for CertCheck.} To obtain an executable
implementation of \isa{certcheck}, we instantiated the \isabellehol
formalization with a concrete syntax and semantics for CNF-XOR formulas, and
extracted source code using \isabellehol's Standard ML extraction mechanism.
The extracted implementation is compiled together with user interface code,
e.g., file I/O, parsing, and interfacing with a trusted random bit generator
and CNF-XOR unsatisfiability checking, as
shown in~\rref{fig:workflow}.  The resulting tool is called {\CertCheck}.

\subsection{Extending ApproxMC to ApproxMCCert.}\label{subsec:approxmccert}
To demonstrate the feasibility of building a (partial) certificate generation
tool, we modified the mainline implementation of \ApproxMC to
accept and use an externally generated source of random bits. We also modified
it to write its internally calculated values of \isa{m} and a log of the
respective models reported by its internal solver to a file.  The resulting
tool is called \ApproxMCCert.
An implementation of {\ApproxMC} (and thus {\ApproxMCCert}) requires
logarithmically many solver calls to \emph{find} the correct value of \isa{m}
and it can employ many search
strategies~\cite{DBLP:conf/ijcai/ChakrabortyMV16}. The partial certificate
format is agnostic to how \isa{m} is found, requiring certification only for
the final value of \isa{m} in each round.

\begin{remark}
It is worth remarking that  {\CertCheck} requires
checking the validity of
$\mathcal{O}(\varepsilon^{-2} \cdot \log \delta^{-1} )$ solutions (each
of size $n$, the number of variables),
and unsatisfiability for $\mathcal{O}(\log \delta^{-1} )$ formulas, while
{\ApproxMC} requires
$\mathcal{O}(\varepsilon^{-2} \cdot \log n \cdot \log \delta^{-1} )$ calls to its
underlying solver.  In the next
section, we instantiate \isa{check\_unsat} with a CNF-XOR unsatisfiability
checking pipeline that generates proofs which are checkable by a
verified checker in polynomial time (in the size of the proofs).
\end{remark}

\subsection{CNF-XOR Unsatisfiability Checking}\label{subsec:cnfxorunsat}

A crucial aspect of {\CertCheck} is its reliance on an external checker for
unsatisfiability of CNF-XOR formulas.  As mentioned in~\rref{sec:related},
there are several prior approaches for certified CNF-XOR reasoning that can be
plugged into {\CertCheck}.

We opted to build our own \emph{native} extension of
FRAT~\cite{DBLP:journals/lmcs/BaekCH22} because none of the previous options
scaled to the level of efficient XOR proof checking needed for certifying
\ApproxMC (as evidenced later in~\rref{sec:experiments}).
For brevity, the new input and proof format(s) are illustrated with inline comments
in~\rref{fig:fratxor}. We defer a format specification to the tool repository.

In a nutshell, when given an input CNF-XOR formula, \CryptoMiniSat has been
improved to emit an unsatisfiability proof in our extended FRAT-XOR format.
Then, our \FRATxor tool elaborates the proof into XLRUP, our extension of
Reverse Unit Propagation (RUP) proofs~\cite{DBLP:conf/itp/HeuleHKW17} with XOR
reasoning. The latter format can be checked using \cakexlrup, our formally
verified proof checker.  Such an extension to FRAT was suggested as a
possibility by Baek et al.~\cite{DBLP:journals/lmcs/BaekCH22} and we bear their
claim out in practice.

\begin{figure*}[t]
\renewcommand{\ttdefault}{pcr}

\centering
\begin{minipage}[t]{.33\textwidth}
\centering
{Input CNF-XOR formula}
\begin{Verbatim}[frame=single,commandchars=\\\{\},fontsize=\small]
p cnf 3 4
1 2 0
-1 -2 0
x 1 2 -3 0
-3 0
\end{Verbatim}
{FRAT-XOR proof file}
\begin{Verbatim}[frame=single,commandchars=\\\{\},fontsize=\small]
...
\textbf{o x 1 1 2 -3 0}
\textbf{i x 2 1 2 0 l 1 2 0}
\textbf{a x 3 3 0 l 1 2 0}
\textbf{i 4 3 0 l 3 0}
a 5 0
...
\end{Verbatim}
\end{minipage}
\qquad
\begin{minipage}[t]{.57\textwidth}
\centering
{XLRUP proof file}
\begin{Verbatim}[frame=single,commandchars=\\\{\},fontsize=\small]
// \redc{Add at XOR ID 1,} \bluec{XOR x 1 2 -3 0,}
// \greenc{from the input formula}
\textbf{\greenc{o} \redc{x 1} \bluec{1 2 -3 0}}
// \redc{Add at XOR ID 2,} \bluec{XOR x 1 2 0,}
// \greenc{implied by clauses 1 and 2}
\textbf{\greenc{i} \redc{x 2} \bluec{1 2 0} \greenc{1 2 0}}
// \redc{Add at XOR ID 3,} \bluec{XOR x 3 0,}
// \greenc{implied by XORs 1 and 2}
\textbf{\redc{x 3} \bluec{3 0} \greenc{1 2 0}}
// \redc{Add at clause ID 4,} \bluec{Clause 3 0,}
// \greenc{implied by XOR 3}
\textbf{\greenc{i} \redc{4} \bluec{3 0} \greenc{3 0}}
// Derive empty clause by RUP,
// hints generated by FRAT-xor
5 0 3 4 0
\end{Verbatim}
\end{minipage}
\caption{(top left) A sample input CNF-XOR formula where XOR lines start with
\texttt{x} and indicate the literals that XOR to $1$, e.g., the line
\texttt{x 1 2 -3} represents the XOR constraint $x_1 \oplus x_2 \oplus
\bar{x}_3 = 1$; (bottom left) a FRAT-XOR proof; (right) an XLRUP proof. The
steps in \textbf{bold} indicate newly added XOR reasoning.
Note that the XOR steps are (mostly) syntactically and semantically
unchanged going from FRAT-XOR to XLRUP, so we focus on the latter here. The
meaning of each XLRUP step (analogously for FRAT-XOR) is annotated in
color-coded comments above the respective line.}
\label{fig:fratxor}
\end{figure*}

\subsubsection{Extending \FRATrs to \FRATxor.}
Our \FRATxor tool adds XOR support to
\FRATrs~\cite{DBLP:journals/lmcs/BaekCH22}, an existing tool for checking and
elaborating FRAT proofs. This extension is designed to be
\emph{lightweight}---\FRATxor does not track XORs nor check the correctness of
any XOR-related steps; instead, it defers the job to an underlying verified proof
checker. Our main changes were:
\begin{inparaenum}[(i)]
\item adding parsing support for XORs;
\item ensuring that clauses implied from XORs can be properly used for further
    clausal steps, including automatic elaboration of
    RUP~\cite{DBLP:journals/lmcs/BaekCH22}; and
\item ensuring the clauses used to imply XORs are trimmed from the proof at
    proper points, i.e., after the last usage by either a clausal or XOR step.
\end{inparaenum}

\subsubsection{Extending \cakelpr to \cakexlrup.}
We also modified \cakelpr~\cite{DBLP:journals/sttt/TanHM23}, a verified proof
checker for CNF unsatisfiability, to support reasoning over XOR constraints.
The new tool supports: \begin{inparaenum}[(i)] \item clause-to-clause reasoning
via RUP steps; \item deriving new XORs by adding together XORs; \item
XOR-to-clause and clause-to-XOR implications.  \end{inparaenum} The main
challenge here was to represent XORs efficiently using byte-level
representations to take advantage of native machine instructions in
XOR addition steps. The final verified correctness theorem for \cakexlrup is
similar to that of \cakelpr~\cite{DBLP:journals/sttt/TanHM23} (omitted here). 

\subsubsection{Modifications to CryptoMiniSat.}
A refactoring of \CryptoMiniSat was performed in response to the bug described
in~\rref{sec:introduction} and in order to add FRAT-XOR proof logging.  As part
of this rewrite, a new XOR constraint propagation engine has been added
that had been removed as part of
BIRD~\cite{DBLP:conf/aaai/SoosM19}---that system did not need it, as it kept
all XOR constraints also in a blasted form.
Furthermore, XOR constraints have been given IDs instead
of a pointer to a \Tbuddy BDD previously used, and all XOR manipulations
such as XOR-ing together XOR constraints, constant folding~\cite{wegman1991constant}, satisfied XOR constraint deletion, etc.,
had to be documented in the emitted FRAT-XOR proof log. Further, \CryptoMiniSat
had to be modified to track which clause IDs were responsible for
recovered XOR constraints. To make sure our changes were correct, we modified
\CryptoMiniSat's fuzzing pipeline to include XOR constraint-generating problems and to
check the generated proofs using our certification tools.

\section{Experimental Evaluation} \label{sec:experiments}

To evaluate the practicality of partial certificate generation
({\ApproxMCCert}) and certificate checking ({\CertCheck}), we conducted an
extensive evaluation over a publicly available benchmark set~\cite{MS20} of
1896 problem instances that were used in previous evaluations of
{\ApproxMC}~\cite{DBLP:conf/cav/SoosGM20,SM22}.
The benchmark set consists of (projected) model counting problems arising from
applications such as probabilistic reasoning, plan recognition, DQMR networks,
ISCAS89 combinatorial circuits, quantified information flow, program synthesis,
functional synthesis, and logistics.
Most instances are satisfiable with large model counts and only
approximately 6\% are unsatisfiable for testing corner cases.

To demonstrate the effectiveness of our new CNF-XOR unsatisfiability checking
pipeline, we also compared it to the three prior state-of-the-art
approaches discussed in~\rref{sec:related}. The approaches are labeled as
follows:
\begin{itemize}
\item[{\cmscake}.] Our new (default) pipeline based on FRAT-XOR (\rref{subsec:cnfxorunsat}); here, \texttt{CMS} is short for \CryptoMiniSat.
\item[{\cmsbuddy}.] The pipeline consisting of \CryptoMiniSat with \Tbuddy, \FRATrs, and a verified CNF proof checker (\rref{sec:related}, item~\rref{item:bdd}).
\item[{\minisatxor}.] The pipeline consisting of \MiniSat with XOR engine, \veripb, and its verified proof checker (\rref{sec:related}, item~\rref{item:pbp})
\item[{\CadCake}.] A state-of-the-art SAT solver \CaDiCaL~\cite{BiereFazekasFleuryHeisinger-SAT-Competition-2020-solvers,DBLP:conf/sat/PollittFB23} which generates proofs checkable by a verified CNF proof checker (\rref{sec:related}, item~\rref{item:cdcl}).
\end{itemize}

We experimented with each of these approaches as the CNF-XOR
unsatisfiability checking pipeline for {\CertCheck}, checking the same suite of
certificates produced by {\ApproxMCCert}.

The empirical evaluation was conducted on a high-performance computer cluster
where every node consists of an AMD EPYC-Milan processor featuring $2\times64$
real cores and 512GB of RAM.  For each instance and tool ({\ApproxMC},
{\ApproxMCCert}, or {\CertCheck}), we set a timeout of 5000 seconds, memory
limit of 16GB, and we used the default values of $\delta=0.2$ and
$\varepsilon=0.8$ for all tools following previous
experimental conventions~\cite{DBLP:conf/cav/SoosGM20}.
For each given tool, we report the PAR-2 score which is commonly used in the
SAT competition.  It is calculated as the average of all runtimes for
solved/certified instances out of the relevant instances for that tool, with
unsolved/uncertified instances counting for double the time limit (i.e., 10000
seconds).

Our empirical evaluation sought to answer the following questions:
\begin{description}
  \item[RQ1] How does the performance of {\ApproxMCGen} and {\CertCheck} compare to that of {\ApproxMC}?
  \item[RQ2] How does the performance of {\cmscake} compare to prior state-of-the-art approaches for CNF-XOR UNSAT checking for use in {\CertCheck}?
\end{description}

\subsubsection{RQ1  Feasibility of Certificate Generation and Checking.}
\begin{figure}[htb]
	\centering
	\includegraphics[scale=0.78]{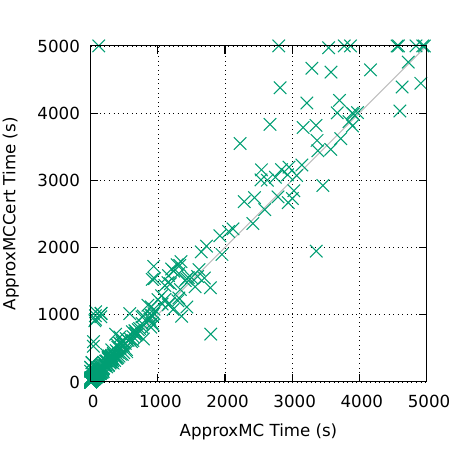}
	\caption{Per instance runtime (s) comparison for {\ApproxMCGen} and {\ApproxMC}.}
	\label{fig:cert-gen}
\end{figure}

We present the results for {\ApproxMC}, {\ApproxMCGen}, and  {\CertCheck} in
Table~\ref{tab:num}.  For certificate generation, our main observation is that
{\ApproxMCGen} is able to solve and generate certificates for $99.3\%$ (i.e.,
1202 out of 1211) instances that {\ApproxMC} can solve alone, and their PAR-2
scores (out of 1896 instances) are similar.  Indeed, in the per-instance
scatter plot of {\ApproxMC} and {\ApproxMCGen} runtimes in~\rref{fig:cert-gen},
we see that for almost all instances, the overhead of certificate generation in
{\ApproxMCGen} is fairly small.  This is compelling evidence for the
practicality of adopting \emph{certificate generation} for approximate counters
with our approach.

\begin{table}[htb]
	\caption{Performance comparison of {\ApproxMC}, {\ApproxMCGen}, and {\CertCheck}. The PAR-2 score is calculated out of 1896 instances for {\ApproxMC} and {\ApproxMCGen}, and out of the 1202 instances with certificates for {\CertCheck}.}
  \centering
	\begin{tabular}{c@{\hskip 0.2in} c@{\hskip 0.2in} c@{\hskip 0.1in}|@{\hskip 0.1in} c@{\hskip 0.2in}}
		\toprule
		& {\ApproxMC}& {\ApproxMCGen} & {\CertCheck} \\
		\hline
		 Counted Instances & 1211 & 1202 & 1018  \\
		PAR-2 Score & 3769 & 3815 & 1743 \\
		\bottomrule
	\end{tabular}
	\label{tab:num}
\end{table}

Turning to the feasibility of certificate checking, we observe in
Table~\ref{tab:num} that {\CertCheck} is able to fully certify $84.7\%$
of the instances (i.e., 1018 out of 1202) with certificates. Of the
remaining instances, {\CertCheck} timed out for 46 and ran out of
memory for 138 instances (no certificate errors were reported in our
latest versions of the tools).  On average, {\CertCheck} requires 4.6
times the runtime of {\ApproxMCCert} across all certified instances.
Note that each instance of {\CertCheck} requires nine separate calls to
the CNF-XOR unsatisfiability checking pipeline (because $\delta =
0.2$).
It is worth emphasizing that in other certificate checking setups, such as the
SAT competitions, one would typically provide an order of magnitude more time
and memory to the checkers compared to solvers. Thus, {\CertCheck} performs
well even though our time and memory limits are stringent. Furthermore, we
believe that {\CertCheck}'s ability to achieve a fairly low PAR-2 score
(computed out of 1202 instances) is compelling evidence for the practicality of
certificate checking in approximate counting.
Future work could explore \emph{parallelized} certificate checking since
each round used in {\CertCheck} can be checked independently of each other.

\subsubsection{RQ2 Comparison of CNF-XOR Unsatisfiability Checkers.}
We present results using various alternative unsatisfiability checking
pipelines as part of {\CertCheck} in~\rref{tab:cert-num}.  Here, we observe
that the use of {\cmscake} allows {\CertCheck} to fully certify significantly more instances than can be certified by prior
approaches, and with a much lower PAR-2 score.
\begin{table}
	\caption{Performance comparison of CNF-XOR unsatisfiability checkers in {\CertCheck}. The PAR-2 score is calculated out of the 1202 instances with certificates for all checkers.}
	\centering
  \begin{tabular}{c@{\hskip 0.1in} c@{\hskip 0.1in}  c@{\hskip 0.1in} c@{\hskip 0.1in} c@{\hskip 0.1in}}
      \toprule
      Total	      &  {\CadCake}	& {\minisatxor} & \cmsbuddy & {\cmscake}	\\
      \midrule
      Counted Instances    &     527     &    563 		& 	623 	&  \textbf{1018}    \\
      PAR-2 Score &     5742    &    5659 		&  	5027 	& \textbf{1743}    \\
      \bottomrule
    \end{tabular}
	\label{tab:cert-num}
\end{table}

\begin{figure}[t]
	\centering
	\includegraphics[scale=0.78]{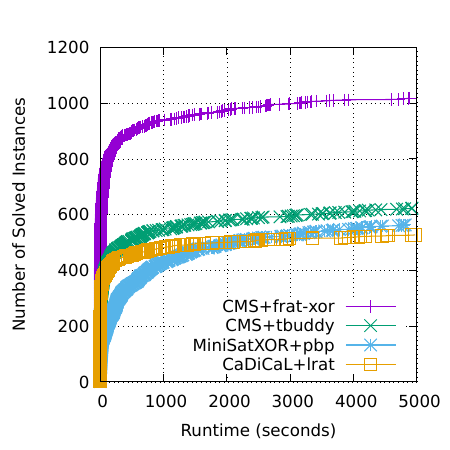}
	\includegraphics[scale=0.78]{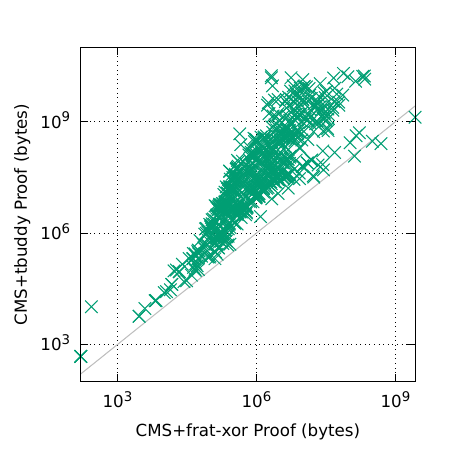}
	\caption{(left) Runtime performance comparison between CNF-XOR unsatisfiability checkers. (right) Per instance CNF-XOR unsatisfiability proof size (bytes) comparison for {\cmscake} and {\cmsbuddy}.}
	\label{fig:runtime}
\end{figure}

Figure~\ref{fig:runtime} (left) visualizes the performance gap between
{\cmscake} and the prior methods using a CDF (cumulative distribution function)
plot; a point $(x,y)$ indicates that the corresponding tool certifies $y$
number of instances when given a timeout of $x$ seconds for each instance.
This plot provides strong justification for our claim of the need to develop
{\cmscake} for native CNF-XOR unsatisfiability proof checking
in~\rref{subsec:cnfxorunsat}.
The ability to log XOR proof steps compactly in our new CNF-XOR
unsatisfiability proof format is also significant.  This is illustrated in
Figure~\ref{fig:runtime} (right) which gives a scatter plot comparing FRAT
(resp.~FRAT-XOR) proof sizes generated by {\cmsbuddy} (resp.~{\cmscake}) within
600 seconds on instances that were successfully certified by {\cmsbuddy}.
Recall that the solver in {\cmsbuddy} supports XOR reasoning and uses {\Tbuddy}
to emit its proof log in terms of a clausal proof system, i.e., without native
XOR proof steps.  Overall, our new proof format achieves an average 30-fold
reduction in proof size, with the maximum reduction reaching up to 8,251 times.

\section{Conclusion and Future Work}
This work shows that it is feasible to use proof assistants to formalize
practical randomized automated reasoning algorithms. Such formalizations are
valuable---our end-to-end certification approach for \ApproxMCCert has led to bug-fixes for both
\ApproxMC and its underlying \CryptoMiniSat solver.

An interesting line of future work would be to support recently proposed
techniques such as \emph{sparse hashing}~\cite{DBLP:conf/lics/MeelA20} or
\textit{rounding}~\cite{DBLP:conf/cav/YangM23} in the context of {\ApproxMC}.
Furthermore, this work leaves preprocessing techniques, such as independent
support identification, out of scope. It is worth noting that efficient
identification of the independent support set, in conjunction with a new
rounding-based algorithm~\cite{DBLP:conf/cav/YangM23}, significantly boosts the
counting performance of {\ApproxMC}; in the experimental setting
of~\rref{tab:num}, this combination solves $1787$ instances with a PAR-2 score
of $625$. Thus, certifying these extensions is a tantalizing avenue for future
research. Another potential line of future work involves developing extensions
for theories other than CNF-XOR model counting~\cite{DBLP:conf/cp/YangM21}.

\subsubsection{\ackname}

  This work has been financially supported by
  the Swedish Research Council grant \mbox{2021-05165},
  National Research Foundation Singapore under its NRF Fellowship Programme [NRF-NRFFAI1-2019-0004], Ministry of Education Singapore Tier 2 Grant [MOE-T2EP20121-0011], Ministry of Education Singapore Tier 1 Grant [R-252-000-B59-114],
  and by A*STAR, Singapore.
  The computational experiments were performed on resources of the National Supercomputing Centre, Singapore \href{https://www.nscc.sg}{https://www.nscc.sg}.
  Part of this work was carried out while some of the authors participated in
  the Spring 2023 \emph{Extended Reunion: Satisfiability} program at the Simons
  Institute for the Theory of Computing and at Dagstuhl workshop 22411
  \emph{Theory and Practice of SAT and Combinatorial Solving}.

\bibliographystyle{splncs04}
\bibliography{paper}

\iflongversion
\newpage
\appendix
\section{Proof of 3-Universality for XOR-based Hash Families}
\label{app:threeunivproof}

An XOR constraint $X$ over a finite set of variables $V$ is a pair $X = (S,b)$ with $S \subseteq V$ and $b \in \{0,1\}$.
For subsets $\omega \subseteq V$, define $X(\omega) = \parity(|\omega \cap S| + b)$.
Intuitively, if $\omega$ is the set of variables assigned to true by a Boolean assignment, then $X(\omega) = 1$ iff the XOR equation $(\bigoplus_{x \in S} x) = b$ is satisfied by that assignment.
Given XOR constraints $X_1, \dots, X_n$, define the corresponding hash function $h : 2^V \to \{0,1\}^n$ where $h(\omega) = (X_1(\omega),\dots,X_n(\omega))$.

For our purposes, XORs are sampled by selecting a subset $S \subseteq V$ and bit $b$ uniformly at random; note that there are $2^{|V|+1}$ possible XORs, each chosen with equal probability.
The XOR-based hash family $H$ is the family of hash funtions $h : 2^V \to \{0,1\}^n$ sampled by randomly choosing $n$ seed XORs independently and uniformly at random and constructing $h$ as above.

Observe that $h(\omega)$ is uniformly distributed in $\{0,1\}^n$ because each output of $X_i(\omega)$ is equiprobable (see also the proof of~\rref{eq:threeunivone} below).
The hash family $H$ is $3$-universal if, in addition, for any three distinct subsets $\omega_1, \omega_2, \omega_3 \subseteq V$, the hash values $h(\omega_1), h(\omega_2), h(\omega_3)$ are independent.
In particular, since $X_1,\dots,X_n$ are sampled independently, it suffices to show that for a random XOR $X$, the output bits $X(\omega_1), X(\omega_2), X(\omega_3)$ are independent.

We prove a slightly more general result.  Let $c_1, c_2, c_3 \in
\{0,1\}$, then randomly sampled XORs over $V$ have the following
independence properties:
\begin{align}
	&\Pr_X \big( X(\omega_1) = c_1 \big)  = \frac{1}{2} \label{eq:threeunivone}\\
	&\Pr_X \big( X(\omega_1) = c_1 \land X(\omega_2) = c_2 \big) = \frac{1}{4} \label{eq:threeunivtwo}\\
	&\Pr_X \big( X(\omega_1) = c_1 \land X(\omega_2) = c_2 \land  X(\omega_3) = c_3 \big) = \frac{1}{8} \label{eq:threeunivthree}
\end{align}

\begin{proof}
	The key idea used throughout is to suitably interchange the sampling order for the XOR constraint.
	In~\rref{eq:threeunivone}, by delaying the sampling of bit $b$, we have:
	\begin{align*}
		\Pr_X \left( X(\omega_1) = c_1 \right) &= \E_{S \subseteq V} \left( \Pr_b (\parity(|\omega_1 \cap S| + b) = c_1 ) \right)\\
		&= \E_{S \subseteq V} \left( \Pr_b (b = \parity(|\omega_1 \cap S| + c_1) ) \right)\\
		&= \E_{S \subseteq V} \left( \frac{1}{2} \right) = \frac{1}{2}
	\end{align*}

	For~\rref{eq:threeunivtwo}, since $\omega_1, \omega_2$ are distinct sets, without loss of generality, there is a variable $x \in \omega_1, x \notin \omega_2$.
	We delay the sampling for $x$, i.e., we first sample an XOR over variables $V - \{x\}$ and then a random bit $b_x$ determining whether variable $x$ is chosen.
	Note that $x \notin \omega_2$ implies the value of $X(\omega_2)$ is independent of the random (delayed) choice of $b_x$. Formally:

	\begin{align*}
		&\Pr_X \left( X(\omega_1) = c_1 \land X(\omega_2) = c_2 \right) \\
		&= \E_{S \subseteq V -\{x\},b} \left( \Pr_{b_x} (\parity(|\omega_1 \cap S| + b_x + b) = c_1 \land \parity(|\omega_2 \cap S| + b) = c_2) \right)\\
		&= \E_{S \subseteq V -\{x\},b} \left( \Pr_{b_x} (\parity(|\omega_1 \cap S| + b_x + b) = c_1) \cdot \indicator(\parity(|\omega_2 \cap S| + b) = c_2)  \right)\\
		&= \E_{S \subseteq V -\{x\},b} \left( \frac{1}{2} \cdot \indicator(\parity(|\omega_2 \cap S| + b) = c_2)  \right)\\
		&= \frac{1}{2} \cdot \E_{S \subseteq V -\{x\},b} \left( \indicator(\parity(|\omega_2 \cap S| + b) = c_2)  \right) = \frac{1}{2} \cdot \frac{1}{2} = \frac{1}{4}
	\end{align*}

	The proof for~\rref{eq:threeunivthree} generalizes the above ideas to a three-way case analysis on the intersections between $\omega_1,\omega_2, \omega_3$.

	\paragraph{Case 1.} There exists $x \in \omega_i$, $x \notin \omega_j$, $x \notin \omega_k$ for distinct $i,j,k \in \{1,2,3\}$.
	Without loss of generality, let $x \in \omega_1$, $x \notin \omega_2$, $x \notin \omega_3$.
	As in the proof for~\rref{eq:threeunivtwo}, we first sample an XOR over variables $S \subseteq V - \{x\}$, which determines $X(\omega_2)$ and $X(\omega_3)$, then a random bit $b_x$ determining whether variable $x$ is chosen.
	The indicator event abbreviated $\dots$ (shown on the last line) corresponds to the pairwise-independence of random XORs from~\rref{eq:threeunivtwo}.
	\begin{align*}
		&\Pr_X \left( X(\omega_1) = c_1 \land X(\omega_2) = c_2 \land X(\omega_3) = c_3\right) \\
		&= \E_{S \subseteq V -\{x\},b} \left( \Pr_{b_x} (\parity(|\omega_1 \cap S| + b_x + b) = c_1) \cdot \indicator(\dots)  \right)\\
		&= \frac{1}{2} \cdot \E_{S \subseteq V -\{x\},b} \left(  \indicator(\parity(|\omega_2 \cap S| + b) = c_2 \land \parity(|\omega_3 \cap S| + b) = c_3) \right) = \frac{1}{8}
	\end{align*}

	For the remaining cases, define four non-overlapping sets of variables $A,B,C,I$ as in the following Venn diagram.
	Note that the unlabeled parts of the diagram are empty, as Case 1 would apply otherwise.

	\begin{center}
		\begin{tikzpicture} [set/.style = {draw,
				circle,
				minimum size = 2cm,
				fill=white,
				opacity = 0.4,
				text opacity = 1}]

			\node (A) [set,label={[label distance=-2.7cm]0:$\omega_1$}] {};
			\node (B) at (60:1.2cm) [set,label={[label distance=-2.5cm]-90:$\omega_2$}] {};
			\node (C) at (0:1.2cm) [set,label={[label distance=0cm]0:$\omega_3$}] {};

			\node at (barycentric cs:A=1,B=1) [left] {$A$};
			\node at (barycentric cs:A=1,C=1) [below] {$C$};
			\node at (barycentric cs:B=1,C=1) [right] {$B$};
			\node at (barycentric cs:A=1,B=1,C=1) [] {$I$};
		\end{tikzpicture}
	\end{center}

	For a subset $S \subseteq V$, we will use $S_A, S_B, S_C, S_I$ to denote its intersection with $A,B,C,I$, respectively.
	Note that the value of each $X(\omega_i)$ is completely determined by $\parity(|S_A|)$, $\parity(|S_B|)$, $\parity(|S_C|)$, and $\parity(|S_I| + b)$, e.g.:
		\begin{align*}
			X(\omega_1) &= \parity(|\omega_1 \cap S| + b) \\
      &= \parity\big(\parity(|S_A|) + \parity(|S_C|) + \parity(|S_I| + b) \big)
	\end{align*}

	Fixing the value of one of the cardinality parities, say $\parity(|S_A|)$, yields exactly one valuation for $\parity(|S_B|)$, $\parity(|S_C|)$, and $\parity(|S_I| + b)$ that satisfy the three equations $ X(\omega_1) = c_1$, $X(\omega_2) = c_2$, $X(\omega_3) = c_3$.
	In other words, there are bits $b_B^i,b_C^i,b_I^i$ for $i = 0, 1$, corresponding to each choice of $\parity(|S_A|) = i$, where
		\begin{align*}
			&\Pr_X \big( X(\omega_1) = c_1 \land X(\omega_2) = c_2 \land  X(\omega_3) = c_3 \big) \\
			&= \sum_i \Pr_X \left(
        \begin{aligned}
        &\parity(|S_A|) = i \land \parity(|S_B|) = b_B^i\\
        &\land \parity(|S_C|)  = b_C^i\land \parity(|S_I| + b) = b_I^i
			\end{aligned}\right) \\
			&= \sum_i \left(
			\begin{aligned}
				&\Pr_{S_A\subseteq A} \big(  \parity(|S_A|) = i \big) \cdot \Pr_{S_B\subseteq B} \big( \parity(|S_B|) = b_B^i \big) \\
				&\cdot \Pr_{S_C \subseteq C} \big(\parity(|S_C|)  = b_C^i \big) \cdot \Pr_{S_I \subseteq I, b} \big(\parity(|S_I| + b) = b_I^i\big)
			\end{aligned}\right) \\
			&= \frac{1}{2} \cdot \sum_i \left(
			\begin{aligned}
			&\Pr_{S_A\subseteq A} \big(  \parity(|S_A|) = i \big) \cdot \Pr_{S_B\subseteq B} \big( \parity(|S_B|) = b_B^i \big) \\
      & \cdot \Pr_{S_C \subseteq C} \big(\parity(|S_C|)  = b_C^i \big)
			\end{aligned}
      \right)
	\end{align*}

	The penultimate step follows from the fact that $A,B,C,I$ are disjoint sets so the respective events involving those sets are independent. The last step follows from delayed sampling of bit $b$.

	\paragraph{Case 2.} Sets $A,B,C$ are all non-empty.
	By a delayed sampling argument on the element in each of these sets:
		\begin{align*}
			&\Pr_X \big( X(\omega_1) = c_1 \land X(\omega_2) = c_2 \land  X(\omega_3) = c_3 \big) \\
			&= \frac{1}{2} \cdot \sum_i \left(
			\begin{aligned}
			&\Pr_{S_A\subseteq A} \big(  \parity(|S_A|) = i \big) \cdot \Pr_{S_B\subseteq B} \big( \parity(|S_B|) = b_B^i \big) \\
      &\cdot \Pr_{S_C \subseteq C} \big(\parity(|S_C|)  = b_C^i \big)
			\end{aligned}
      \right) \\
			&= \frac{1}{2} \cdot \sum_i \left(
			\frac{1}{2} \cdot \frac{1}{2} \cdot \frac{1}{2}\right) = \frac{1}{8}
	\end{align*}

	\paragraph{Case 3.} Exactly one of the sets $A,B,C$ is empty.
	Without loss of generality, let $A = \emptyset$, so $\parity(|S_A|) = 0$, which forces $i = 0$. By delayed sampling for the element in each of the non-empty sets $B,C$:
		\begin{align*}
			&\Pr_X \big( X(\omega_1) = c_1 \land X(\omega_2) = c_2 \land  X(\omega_3) = c_3 \big) \\
			&= \frac{1}{2} \cdot \sum_i \left(
			\begin{aligned}
			&\Pr_{S_A\subseteq A} \big(  \parity(|S_A|) = i \big) \cdot \Pr_{S_B\subseteq B} \big( \parity(|S_B|) = b_B^i \big) \\
      &\cdot \Pr_{S_C \subseteq C} \big(\parity(|S_C|)  = b_C^i \big)
			\end{aligned}
      \right) \\
			&= \frac{1}{2} \cdot \Pr_{S_B\subseteq B} \big( \parity(|S_B|) = b_B^0 \big) \cdot \Pr_{S_C \subseteq C} \big(\parity(|S_C|)  = b_C^0 \big)\\
			&= \frac{1}{2} \cdot \frac{1}{2} \cdot \frac{1}{2} = \frac{1}{8}
	\end{align*}

	No other cases are possible because $\omega_1, \omega_2, \omega_3$ are distinct sets.\qed
\end{proof}

\fi

\end{document}